\UseRawInputEncoding
\documentclass[aps,prx,showpacs,twocolumn,groupedaddress]{revtex4-1}

\usepackage{times,xspace}
\usepackage{graphicx,graphics,color,epsfig}
\usepackage{amsmath}
\usepackage{amssymb}
\usepackage{amsfonts}
\usepackage{amsfonts}
\usepackage{epstopdf}
\usepackage{bm}
\usepackage{hyperref}
\usepackage{color}
\usepackage{float}
\usepackage[normalem]{ulem}

\def\be{\begin{equation}}
\def\ee{\end{equation}}
\def\ba{\begin{eqnarray}}
\def\ea{\end{eqnarray}}

\newcommand{\figref}[1]{Fig.~\ref{#1}}

\begin{document}

\title{A hierarchy of multi-moment skyrmion phases in twisted magnetic bilayers}

\author{Sujay Ray, and Tanmoy Das}
\email{tnmydas@iisc.ac.in}
\affiliation{Department of Physics, Indian Institute of Science, Bangalore-560012, India}

\date{\today}
 
\begin{abstract}
The recent discovery of two-dimensional (2D) Van der Waals (VdW) magnets is a crucial turning point in the quantum magnet research field, since quantum fluctuations and experimental difficulties often elude stable magnetic orders in 2D. This opens new doors to delve for novel quantum and topological spin configurations, which may or may not have direct analogs in bulk counterparts. Here we study a twisted bilayer geometry of 2D magnets in which long-range spin-spin interactions naturally commence along the inter-layer Heisenberg ($J_{\perp}$) and dipole-dipole ($J_{\rm D}$) channels. The $J_{\perp}-J_{\rm D}$ parameter space unveils a hierarchy of distinct skyrmions phases, ranging from point-, rod-, and ring-shaped topological charge distributions. Furthermore, we predict a novel topological antiferroelectric phase, where oppositely-charged antiskyrmion pairs are formed, and the corresponding topological dipole moments become ordered in a N\'eel-like state $-$ hence dubbed topological antiferroelectric state. The results  indicate that twisted magnetic layer provides a new setting to engineer and tune a plethora of novel and exotic skyrmion phases and their dynamics.
\end{abstract}
\vspace{0.5in}

\maketitle

\section{Introduction}

Skyrmion is a particle dual to a topological configuration of the $O(3)$ fields (read spin) in a 2+1 dimension\cite{Skyrme,SkyrmionGenealTheory,SkyrmionReview1,Skyrmionbook}. Such a spin configuration is an allowed classical solution of the non-linear sigma model. However, the main challenge lies in stabilizing a skyrmion solution at a saddle-point energy minimum, requiring distinct magnetic interactions and frustration. Long-range dipole-dipole interaction, in addition to easy-plane magnetic anisotropy and magnetic field, was initially proposed to mediate skyrmion solution.\cite{SkyrmionDDI}
Dzyaloshinskii-Moriya interaction (DMI) brings in the chiral spin-spin interaction required for a skyrmion solution\cite{SkyrmionGenealTheory,SkyrmionDMI,skDMI1,skDMI2,skDMI3,skDMI4}. Competition between DMI and ferromagnetic exchange interactions makes a spatially varying magnetic texture to have a lower energy than the mean-field long-range magnetic order. Geometrical and magnetic frustration can also stabilize skyrmion structures\cite{magnetic_frus1,magnetic_frus2,magnetic_frus3,SkyrmionFrus}. This is induced, for example, in a triangular lattice by the competition between a ferromagnetic nearest neighbour (NN) exchange with an  antiferromagnetic next nearest neighbour (NNN) interaction. Apart from these, more recently, spin-orbit coupling\cite{SkyrmionSOC}, Kondo coupling\cite{magnetic_frus4}, and magnetic disorder with the application of magnetic pulse \cite{skyrmion_disorder} are shown to assist skyrmion solution. Proposals to obtain skyrmions via quantum Hall substrates\cite{SkyrmionQHI} and optical lattices\cite{SkyrmionBEC} are also presented.

By now, there have been several materials realizations of the skyrmions, mainly in systems with DMI, such as Bloch-type skyrmions in MnSi\cite{MnSi}, Co$_{0.5}$Fe$_{0.5}$Si,\cite{FeCoSi,Tokura10,MildeKohlerSci13} Cu$_2$OSeO$_3$,\cite{Cu2OSeO3} CoZnMn,\cite{CoZnMn} and FeGe\cite{FeGe}, and  N\'eel-type skyrmions 
in ferromagnetic heavy-metals\cite{heavymetals}, Kagome lattice Fe$_3$Sn$_2$\cite{Kagome}, magnetic films,\cite{WisendangerReview}, and antiskyrmions in Mn$_2$RhSn\cite{Mn2RhSnParkin,Antiskyrmion}. Experimental abilities to breed,\cite{MildeKohlerSci13,WritingSkyrmion} mobilize,\cite{MobileSkyrmion} rotate\cite{MochizukiNatMat} skyrmions are also demonstrated recently.\cite{Hofman,WisendangerReview}  
Owing to the topological robustness, skyrmions have numerous potential applications in quantum information,\cite{SkyrmionReviewTech} racetrack memory\cite{RacetrackParkin,Racetrack2}, which demand enhanced materials flexibilities and tunabilities. 

Obtaining skyrmion solutions in 2D systems will give new opportunities for science and applications. Recently, long-range magnetic order has been observed in 2D Van der Waals (VdW) chalcogenides, halides, and related materials. Intrinsic antiferromagnetic order is observed in monolayer FePS$_{3}$, \cite{FePS3,FePS3a}, and in $M$P$X_3$($M$=Mn,Fe,Co,Ni; $X$=S,Se)\cite{MPX,MPXa}. Later on, many VdW materials such as Fe$_3$GeTe$_2$\cite{FGT,FGTa}, Mn$X_2$($X$=S,Se)\cite{MnX,MnXa}, V$X_2$($X$=S,Se,Te)\cite{VX,VXa,VXb} were found to be intrinsic ferromagnet. Another exciting family of 2D vdW magnets is the Cr based materials Cr$X_3$ ($X=$I,Br,Cl)\cite{CrX,CrXa}, which are ferromagnets in monolayer, but antiferromagnets in bilayer structure, and the two orderings are externally tunable.\cite{CrXbi,CrXbia} Theoretical and experimental efforts to obtain skyrmions in 2D systems are present. Continuum theory of magnetization field in the non-linear sigma model and Landau-Lifshitz-Gilbert in monolayer and bilayer Moir\'e systems show the existence of Ne\'el type skyrmions.\cite{CrI3_DFT,Moire_magnets,skTBG,Substrate,skFerroDMI} Recently, a skyrmion phase is observed in 2D Fe$_3$GeTe$_2$ on (Co/Pd)$_n$ superlattice\cite{FGT_CoPd} and Fe$_3$GeTe$_2$/$h$-BN heterostructure\cite{FGT_hBN} due to their sizable DMI strength.

To strategize a new mechanism of the skyrmion, it is worth revisiting it's key ingredients. Firstly, topological skyrmion configurations in 2+1 dimension generally belong to the  homotopy group $\pi_2(\mathbb{S}^2)\cong \mathbb{Z}$. The homotopy mapping is exact when both the coordinate space and the target (spin) space are compact $\mathbb{S}^2$. The constraint $|{\bf S}|=1$ compactifies the spin space. When a spin configuration has a one-to-one correspondence with the spatial dimensions, this in turn compactifies the position space $\mathbb{R}^2\rightarrow \mathbb{S}^2$. The resulting one-to-one mapping guarantees the spin configuration to be topological with its skyrmion charge $Q \in \mathbb{Z}-$ an integer winding number. We can reverse the above reasoning for a bottom-up approach. If the effective magnetic field ${\bf B}({\bf r})$, experienced by a local spin due to the surrounding spins and extrinsic fields, lives on a Bloch sphere $\mathbb{S}^2$, then within the minimal Zeeman-like coupling, the field would lay the ground for a topological configuration for the spins. The second essential requirement is that the local field configuration must concomitantly promote a saddle-point energy minimum to stabilize a skyrmion structure.

\begin{figure*}[t]
\includegraphics[width=0.95\textwidth]{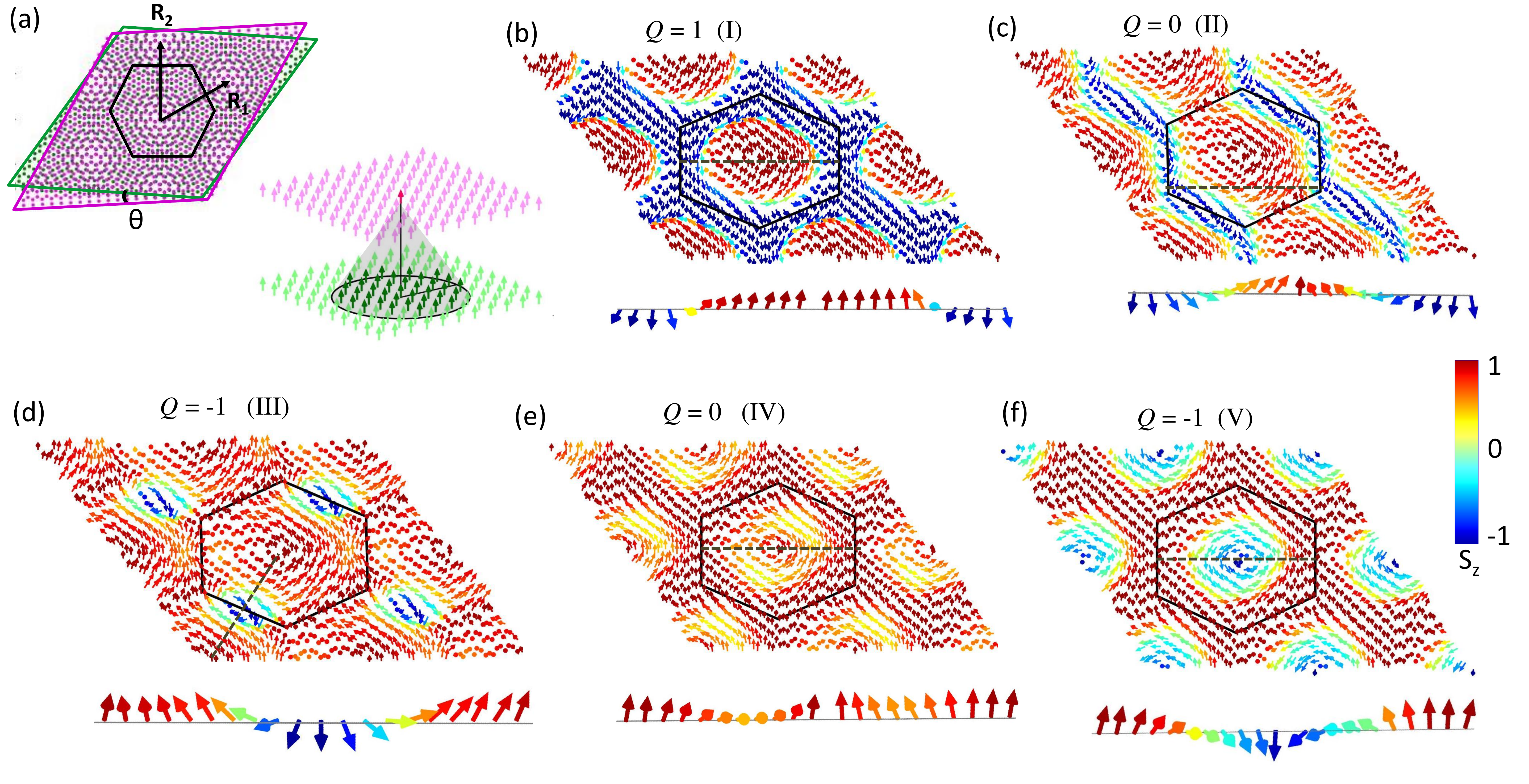}
\caption{{\bf Twisted bilayer setup and five distinct topological spin configurations in real space}. (a) Schematic diagram of twisted bilayer honeycomb ($\theta$ = twist angle) lattice with a Moir\'e pattern is shown along with two lattice translation vectors ${\bf {R}}_{1,2}$. A 3D illustration is shown on the lower right corner, in which a single spin of the upper layer (bright magenta color) interacts with bottom-layer spins within a circular region (dark green color) of radius r$_{\rm cut}$. (b-f) Five distinct spin textures (enumerated by Phase-I to Phase-V) obtained in the $J_{\perp}-J_{\rm D}$ parameter spare (see Fig.~\ref{fig2}). Arrows denote the spin ${\bf s}$ vector, while red to green color-gradient denotes $s_z=1$ to $-1$ values. Dashed thin line in each panel indicates the direction along which a 2D spin projection is shown in the corresponding inset. (b),(d),(f) We show three skyrmion structures with distinct charge density (shown in Fig.~\ref{fig3}) and integer topological charge $Q$. (c) Phase-II corresponds to a novel higher-order topological phase with streamlines of down spins, and topological dipole moments of antiskyrmions pairs (see Fig.~\ref{fig4}). (e) This is a trivial topological phase with finite noncollinear ferromagnetic moment. 
}
\label{fig1}
\end{figure*}

Guided by these principles, and with the recent discoveries of 2D magnets, we lay a blueprint for novel and multifaceted skyrmions (and antiskyrmions) in twisted magnetic bilayers. We construct a Moir\'e superlattice of spins formed in twisted bilayer of VdW magnetic layers with ferromagnetic order at the bottom and $O(3)$ spin dynamics on the top layer. The setup is illustrated in Fig.~\ref{fig1}(a). We include Heisenberg exchange terms $J_{||}$ (for intra-layer) and $J_{\perp}$ (for inter-layer) interactions, and the inter-layer dipole interaction $J_{\rm D}$ as shown in DFT calculations to be dominant in such setup.\cite{CrI3_DFT} We carry out Monte Carlo simulations to determine the microscopic ground state spin configurations at low-temperature, and sweep the entire $J_{\perp}/J_{||}$ and $J_{D}/J_{||}$ parameter space. We identify three distinct skyrmion phases with topological charges $Q=\pm 1$. whose topological charge distributions reveal a previously unknown hierarchy of point-, rod-, and circle - shape in different topological phases. We predict a novel topological spin configuration in the vicinity of the $J_{\perp}/J_{\rm D} \sim -0.4$. We find that near the Moir\'e lattice sites, a pair of spatially separated and oppositely charged antiskyrmions is formed and govern a topological (`electric') dipole moment. More interestingly, such dipoles are found to align anti-ferroelectrically between the nearest-neighbor sites of the Moir\'e lattice, and produce a N\'eel like order for the topological electric dipole moment. We explain these results with a dual electromagnetic theory, demonstrating the `electric field' lines for all topological charge distributions. We also study the `x-ray diffraction' (XRD) pattern of the topological charges (topological charge-charge correlation function) as well as the spin-spin correlation functions to elucidate the crystallization and phase transitions of the topological charge centers and dipoles. These results expand the list of possible skyrmion and magnetic phases (N\'eel and Bloch phase) obtained in continuum models to a hierarchy of skyrmions and its higher order topological phase.\cite{continum_model, Moire_magnets}

\begin{figure}[t]
\includegraphics[width=0.45\textwidth]{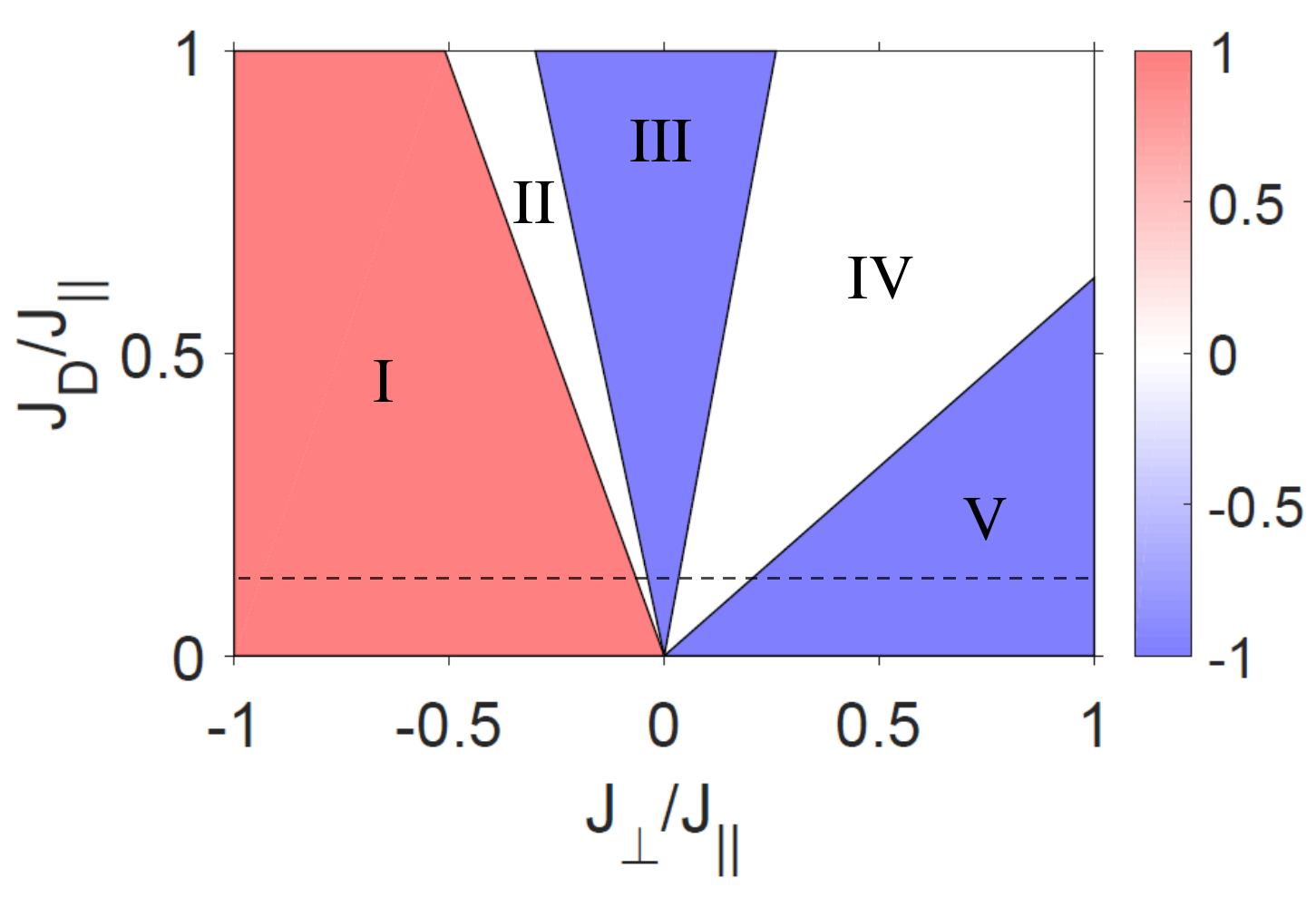}
\caption{{\bf Phase diagram of five distinct topological and quantum phases.} The inter-layer exchange coupling $J_{\perp}$ and dipole interaction $J_{\rm D}$ are varied with respect to the in-plane exchange coupling $J_{||}$. $J_{\rm D}<0$ is an unphysical value, however, mathematically it leaves the phase diagram symmetric when the sign of $J_{\perp}$ is also reversed. Red and blue color distinguishes skyrmion charges of $Q=\pm 1$. Phase-II, although features a net $Q=0$, exhibits a novel topological dipole moment of antiskyrmions which produces an antiferroelectric order state. The horizontal dashed line  indicates the realistic parameter range as deduced in a DFT calculation.\cite{CrI3_DFT}
}
\label{fig2}
\end{figure}

\section{Realization}
We consider a single layer honeycomb magnet (magenta color) placed on a single layer magnetic substrate (green color) of the same lattice structure and lattice constant, as shown in Fig.~\ref{fig1}(a). The distance between the layers $d$ is taken to be same as the lattice constant of the honeycomb lattice. The spin in the substrate layer is fixed to be a collinear ferromagnetic state. This can be achieved with a strong bulk ferromagnetic material as studied in the literature\cite{Substrate}. We primarily focus on small relative twist angles $\theta$ which give the commensurate Moir\'e superlattices. We focus on $\theta=1.61^{o}$, which gives a hexagonal Moir\'e lattice with $a=35.6a_0$, where $a_0$ is the lattice constant of the single layer system. A critical number of atoms in a Moir\'e supercell, determined by the twist angle, is important to stabilize a skyrmion. Above this critical value, the obtained topological phase diagram remains essentially invariant to twist angles and number of atoms, except that the skyrmion radius grows with the Moir\'e cell dimension. So, there is an upper critical value of the twist angle ($\sim 2^{0}$) above which the Moir\'e unit cell becomes small enough and the magnetic unit cell is no longer commensurate with the Moir\'e unit cell.

The general computational strategy is as follows. We split the full Hamiltonian into two parts: $H=H_{1}+H_{2}$, where $H_1$ and $H_2$ are the intra-layer, and inter-layer parts. The intra-layer Hamiltonian consists of a nearest-neighbor Heisenberg exchange and a spin asymmetry term, for both layers. The inter-layer term consists of many neighbors Heisenberg exchange term and the dipole-dipole interaction term. We then integrate out the bottom layer's spin to obtain an effective Hamiltonian for the top layer as $H_{\rm top}\sim H_1+{\bf B}\cdot {\bf s}$, where ${\bf B}$ is the effective magnetic field exerted from the bottom layer on the top layers spin ${\bf s}$. The Moir\'e periodicity is imposed by expanding these terms in the plane wave basis of the Moir\'e supercell. We solve $H_{\rm top}$ within the Monte Carlo simulation.

We now give the details of the model. We label the spin variables for the top and bottom (subtrate) layers by ${\bf s}$, and ${\bf S}$, respectively. The 2D VdW systems in single and bilayer setups are observed to show an in-plane ferromagnetic (and out-of-plane ferro- or anti-ferromagnetic) order with the spin quantization axis to be out-of-plane ($z$-direction).\cite{CrI3,Chalcogenides} Such a magnetic ground state is reproduced by the model: 
\begin{eqnarray}
H_{1}({\bf s})=-J_{||}\sum_{<ij>}{\bf s}_{i}.{\bf s}_{j}-K\sum_{i}\left(s_{iz}\right)^{2},
\label{Eq:H1}
\end{eqnarray}
where $i,j$ lattice sites within a Moir\'e supercell. The first term is the nearest neighbor Heisenberg interaction with coupling constant $J_{||}>0$ for a ferromagnetic phase, and $K$ gives the $z$-axis asymmetry, breaking the $O(3)$ spin degeneracy. $H_1({\bf S})$ term gives the corresponding Hamiltonian for the bottom layer, with $J_{||}$ and $K$ kept fixed. 

The inter-layer interaction $H_2$ is the crucial part. Depending on the twist angle, especially at small twist angles, a spin at the top layer interacts with several neighboring bottom layer spins, and hence the inter-layer interaction involves terms beyond the nearest-neighbor exchange interaction. The inter-layer interaction is mainly dominated by several nearest-neighbor exchange interaction $H_{\rm ex}$ and dipole-dipole interaction ($H_{\rm D}$) as $H_2({\bf S},{\bf s}) = H_{\rm ex}+H_{\rm D}$ where
\begin{eqnarray}
H_{\rm ex}&=&-\sum_{ij}J_{\perp}({\bf r}_{ij}){\bf S}_{i}.{\bf s}_{j},
\label{Eq:Hex}\\
H_{\rm D} &=& J_{\rm D}\sum_{ij}\frac{1}{r_{ij}^{3}} \left[{\bf S}_{i}.{\bf s}_{j}-3({\bf S}_{i}.\hat{{\bf r}}_{ij}).({\bf s}_{j}.\hat{{\bf r}}_{ij})\right].
\label{Eq:HD}
\end{eqnarray}
$i$, $j$ are the site indices in the bottom and top layer, respectively. ${\bf r}_{ij}={\bf r}_{i}-{\bf r}_{j}$ with $\hat{{\bf r}}_{ij}$ being the corresponding unit vector. Since both interactions are long-ranged, we need to set a cutoff radius $r_{\rm cut}$, Fig.~\ref{fig1}(a) (inset). Due to higher power of $r_{ij}$ in the denominator, the result converges quickly by $r_{\rm cut}< 20a_0$ which is much smaller than the Moir\'e latiice size $\sim 35a_0$.

Next, we integrate out the bottom layer's spins ${\bf S}$, and define an effective magnetic field at the top layer at ${\bf r}_i$ as ${\bf B}({\bf r}_i)={\bf B}_{\rm ex}({\bf r}_i)+{\bf B}_{\rm D}({\bf r}_i)$, where ${\bf B}_{\rm ex}$, and ${\bf B}_{\rm D}$ distinguish the contributions from the exchange and dipole-dipole interaction terms as ${\bf B}_{\rm ex}({\bf r}_{i})= -\frac{\partial H_{\rm ex}}{\partial{\bf s}_i}
=\frac{J_{\perp}}{2}\sum_{a}e^{i{\bf G}_{a}.{\bf r}_{i}}$, and ${\bf B}_{\rm D}({\bf r}_{i})
=-\frac{\partial H_{\rm D}}{\partial{\bf s}_i}$. We set ${\bf S}_j= {\hat z}$ for all unit spins at the bottom layer. ${\bf G}_a (a=1$ - $6)$ are the six minimal reciprocal lattice vectors of the Moir\'e superlattice. We include the six possible smallest reciprocal lattice vectors in the Moir\'e lattice and the results do not change with the inclusion of negligibly small contribution of higher reciprocal lattice vectors. Dipole interaction is known to be an useful ingredient for (generally bubble-type) skyrmions and antiskyrmions,\cite{SkyrmionDDI,SkyrmionReview1,Antiskyrmion} but are significantly weaker in strength in real materials. In twisted bilayer system, however, the intra-layer dipole interaction is considerably enhanced, and is found here to be detrimental to the bubble or Bloch skyrmion phases (see below), while promoting novel and distinct skyrmion phases.

In this algorithm, the bottom layer's effects can be cast into a Zeeman-like term at the top layer. The full Hamiltonian hence takes the form
\begin{equation}
H=H_1 - \sum_i {\bf B}({\bf r}_i)\cdot {\bf s}_i.
\label{Eq:Hamfinal}
\end{equation}
In this way, the total Hamiltonian simplifies to the form used in the introduction for discussing how the in-plane spin is enforced to lie on a topological compact space $\mathbb{S}^2$ by tailoring the `long-range' Zeeman coupling. In our Monte Carlo simulation we minimize the local Hamiltonian for a single spin with temperature annealing as well as parameter annealing, and check for the total energy convergence of the system in each case. In all our calculations, the minimization of the local Hamiltonian corresponds to minimization of the total energy. Details of our Monte Carlo simulation can be found in the Appendix~B.

Our model and parameters are justified as follows. Note that the dipole-dipole interaction $J_{\rm{D}}$ is only included in the inter-layer Hamiltonian. This is because, in VdW magnets, DFT calculation has indicated that  the intra-layer exchange is almost 10 times larger than the inter-layer exchange\cite{CrI3_DFT}.
Therefore, any small amount of inter-layer dipole-dipole interaction can have considerable effect on the spin configuration, whereas the intra-layer dipole interaction is always subsided by the large exchange interaction.
Furthermore, the DMI in the existing 2D VdW magnets is small compared to strong intra-layer exchange interaction. Therefore, realistic small values of the intra-layer dipole interaction and DMI interaction do not impact our phase diagram (see Discussion section). The realistic parameter range in CrI$_3$ is indicated by horizontal dash line in Fig.~\figref{fig2}, as deduced from a DFT calculation\cite{CrI3_DFT}.
Interestingly, the phase diagram essentially remains the same below and above this parameter range, emphasizing that all the phases are attainable in the existing materials.  

\section{Results}

\begin{figure}[t]
\includegraphics[width=0.45\textwidth]{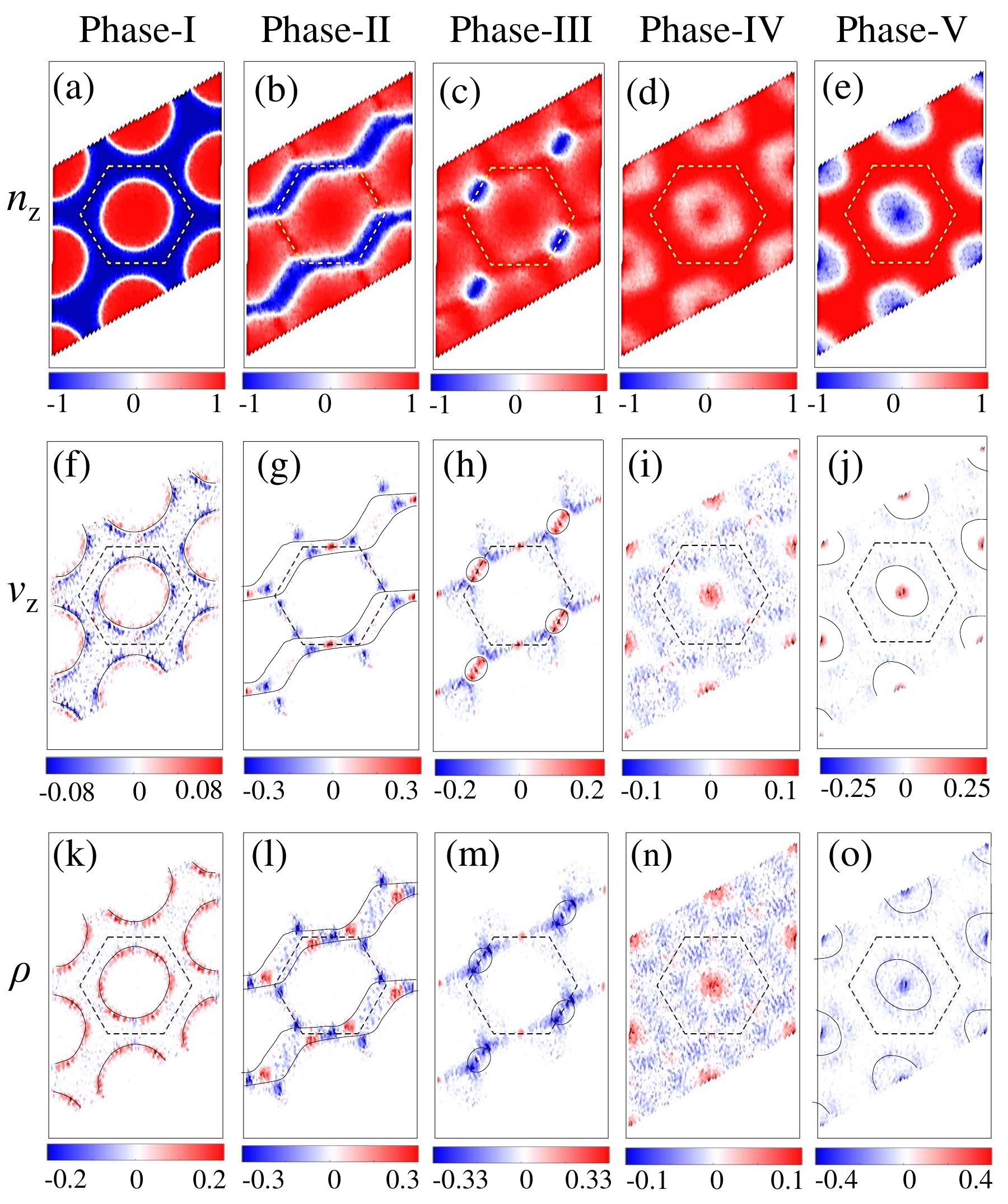}
\caption{{\bf Local variation of polarity, vorticity, and topological charge densities in the five different topological phases.} (a-e) The $z$-component of the polarity density $n_{z}({\bf r})$ (see main text) is plotted in blue to red colormap, denoting down to up spin components of $s_z$. (f-j) We plot the vorticity density $v_z({\bf r})$ with blue to red colormap. The black solid line marks the $n_z=0$ domain wall  boundary. (k-o) Corresponding topological charge density $\rho ({\bf r})$ is plotted here. In Phase-II, IV, although charge centers are formed but the net charge $Q=0$. In Phase-II, fractional charge centers are confined by streamline domain wall, giving a novel topological dipole moment. There is no domain wall of $n_z$ in Phase-IV and hence its a trivial phase. (Different color scales are for the five phases in the middle and bottom panels.) }
\label{fig3}
\end{figure}

\subsection{Phase diagram}
Our simulation yields a plethora of quantum and topological phases; five such distinct configurations are identified in \figref{fig1}(b)-(f). We do not however find any homogeneous mean-field solutions at finite twist angles, while at zero twist angles we only find either a ferromagnetic or an antiferromagnetic order depending on the sign of $J_{||}$. The corresponding phase diagram is presented in \figref{fig2} for $J_{\perp}/J_{||}$ and $J_{\rm D}/J_{||}$ with $J_{||}>0$. In the phase diagram, the red and blue shaded areas denote distinct skyrmion phases with topological invariant $Q=\pm 1$, respectively. The white regions represent a higher-order dipoler antiskyrmion phase (Phase-II), and a trivial phase (Phase-IV). From \figref{fig1}(b),(d), and (f), it is evident that the three skyrmion phases, denoted by Phase-I, III, and V, are characteristically exclusive, which will be distinguished below in multiple ways. Phase-II has zero net topological charge, but possess fractional charge centers of opposite sign, and hence a
topological dipole moment which are antiferroelectrically ordered. The Phase-IV bears no topological or exotic quantum order (except a finite magnetization due to a collinear spin ordering). 

The particle dual of the skyrmion is a topological charge denoted by $Q=\int d^2{\bf r} \rho({\bf r})$, where $\rho({\bf r})$ is the topological charge density. With its corresponding current density $\mathcal{J}_{x,y}$, we concisely define the three-component density operators $\mathcal{J}_{\mu}=(\rho,\mathcal{J}_x,\mathcal{J}_y)$ as
\begin{eqnarray}
\mathcal{J}_{\mu}({\bf r}) = \frac{\epsilon_{\mu\nu\tau}}{8\pi}{\bf n}\cdot\partial_{\nu}{\bf n}\times \partial_{\tau}{\bf n},
\end{eqnarray}
where $\mu,\nu,\tau=0,x,y$ are time-space indices, and the ${\bf r}$ dependence on the unit vector field ${\bf n}={\bf s}/|{\bf s}|$ is implied. We define the vortex density as ${\bf v}({\bf r}) = \epsilon_{\nu\tau}\partial_{\nu}{\bf n}\times \partial_{\tau}{\bf n}$. In our layered geometry and with the $z$-axis asymmetry, it is natural to expect that that the vorticity of the spin-texture commences in the $xy$-plane, i.e, $v_z$ dominates. Then the corresponding polarity density is simply governed by $n_z({\bf r})$. The $z$-components of the polarity density $n_z({\bf r})$, the vortex density $v_z({\bf r})$ and the charge density $\rho({\bf r})$ are investigated in Fig.~\ref{fig3} in three different rows for the five distinct phases (different columns). 

The mechanism of skyrmions and antiskyrmions is retrieved as follows. It is known from the topological band theory\cite{DasRMP} that the polarity field (equivalent to the Dirac mass for fermion fields) forms a nodal (closed) contour, across which $n_z({\bf r})$ changes sign $-$ this is called the domain wall (see top row in Fig.~\ref{fig3}). The vorticity field $v_z({\bf r})$ inside the domain wall acquires singularity $-$ either point- or rod-, or ring- shaped singularity $-$ and cannot be removed with a trivial gauge transformation (see middle row in Fig.~\ref{fig3}). Hence the topological charge density $\rho({\bf r})$ becomes confined within the domain wall (see bottom row in Fig.~\ref{fig3}). The homotopy mapping of the ${\bf n}({\bf r})$ field on $\mathbb{S}^2$ in the ${\bf r}$-space quantizes the topological charge $Q\in \mathbb{Z}$, where the integration of ${\bf r}$ is performed within a single domain wall of the polarity density [see black solid lines in Fig.~\ref{fig3}(f-o)].

\subsection{Skyrmion hierarchy}
Phase-I is a topological phase with $Q=-1$, and is present in most of the $J_{\perp}/J_{||}<0$ region. It is destabilized at a small value of $J_{\perp}$ by stronger dipole interaction $J_{\rm D}$. The distributions of $n_z$, $v_z$, and $\rho$ for Phase-I are shown in the left-most column in Fig.~\ref{fig3}. The polarity density map $n_z({\bf r})$ demarcates a sharp and circular domain wall boundary, which reminds us of a magnetic bubble observed in astronomical space, as well in  magnetic systems.\cite{MagBubble} The vorticity and charge densities of this phase, however, reveal much richer structures unknown before. In Fig.~\ref{fig3}(f), we find that the nodal ring of the polarity density (black line) encloses a circular vortex density $v_z({\bf r}$) structure. In fact, $v_z({\bf r})$ is positive (negative) outside (inside) the domain wall, and share the same nodal ring as that of $n_z({\bf r})$. In what follows, the charge density $\rho({\bf r})$ also acquires a singular ring geometry, confined by the domain wall boundary, see Fig.~\ref{fig3}(k). This phase is also topologically equivalent to the Dirac nodal ring state\cite{DiracNodal} in the electronic structure in which the topology is defined via Berry gauge connection.

As we switch $J_{\perp}\rightarrow -J_{\perp}$, keeping all other interactions fixed, we obtain a characteristically distinct skyrmion texture (denoted by Phase-V) with opposite charge ${\it Q}$. Unlike the sharp domain wall in Phase-I, $n_z$ varies smoothly with ${\bf r}$, and forms a (nearly) elliptical domain wall in Phase-V. Moreover, the Phase-V has a point-like topological charge center sitting at the Moir\'e supercell center. Hence as opposed to ring-singularity in Phase-I, Phase-V acquires point- singularity. The asymmetry between the Phase-I and Phase-V at $\pm J_{\perp}$ for fixed $J_{\rm D}$, results from the competition between $J_{\perp}$ and $J_{\rm D}$. The phase diagram is reversal between $\pm J_{\perp}$ for $J_{\rm D}\rightarrow -J_{\rm D}$.

The skyrmion Phase-III occurs in the vicinity of $J_{\perp}\sim 0$, and is mainly stabilized by the long-range (out-of-plane) dipole-dipole interaction $J_{\rm D}$. The magnetic domains are elliptical in shape, and concentrated at two sides of the Moir\'e supercell. As seen in Fig.~\ref{fig3}(h) and (m), the $n_z$ nodal contour confines a fixed-sign vorticity field (positive), and hence the topological charge distribution (negative since $n_z<0$) becomes quantized. The topological charge density is distributed inside the elliptical domain wall and gives a rod-line singularity. Such a rod-like topological charge distribution repeats periodically [see Fig.~\ref{fig3}(m)] and give a nematic or smectic crystal.

In all three skyrmion phases, each Moir\'e supercell contains a single skyrmion configuration. Therefore, a suitable characteristic length scale associated with different skyrmions can be defined by the domain wall contour ${\bf r}_{\rm d}$ where $n_z({\bf r}_{\rm d})=0$. This condition is very much satisfied where the effective magnetic field due to the bottom layer along the $z$-direction vanishes, i.e., $B^z_{ex}({\bf r}_{\rm d}) + B^z_{D}({\bf r}_{\bf d}) = 0$. From the expression for $B_{\rm ex}$, and $B_{\bf D}$ given in the Appendix~A, we see that ${\bf r}_{\rm d}$ depends on the ratio $J_{\perp}/J_{\rm D}$, and the inter-layer distance $d$ for a given Moir\'e lattice. Depending on the ratio $J_{\perp}/J_{\rm D}$, the condition can turn into an equation of a circle or an ellipse, as we also find numerically. In Phase-III, the domain wall takes an elliptic shape in Fig.~ \ref{fig3}(h) and the two topological charge centers we find here in Fig.~\ref{fig3}(m) sits at the two focal points.

It is then easy to grasp that a skyrmion phase transition occurs when domain wall radius ${\bf r}_{\bf d}$ either shrinks to zero or expands to the Moir\'e cell boundary ${\bf R}_{1,2}$. The phase transition between Phase-I to Phase-II occurs when ${\bf r}_{\rm d}={\bf R}_1$ or ${\bf R}_2$. The rotational symmetry breaking renders a small domain wall to form in Phase-III with opposite polarity at one of the Moir\'e supercell's site for small values of $J_{\perp}$, as seen in Fig.~\ref{fig3}(c). This small domain wall then shrinks to zero with the sign reversal of $J_{\perp}$ which disfavors the domain wall of negative polarity. Curiously, there still exists a finite vortex structure and a finite charge density in the trivial Phase-IV. But owing to the absence of a compact domain wall of the polarity density, the net topological winding number vanishes. Hence, the Phase-IV corresponds to a quasi-uniform ferromagnertic (or antiferromagnetic phase if $J_{||}<0$) as seen in the untwisted CrI$_3$ bilayer samples.\cite{CrI3_DFT} Finally, large $J_{\perp}$ creates another compact domain wall at the center of the Moir\'e supercell in which the topological charge is $Q=-1$.

\begin{figure}[t]
\includegraphics[width=0.45\textwidth]{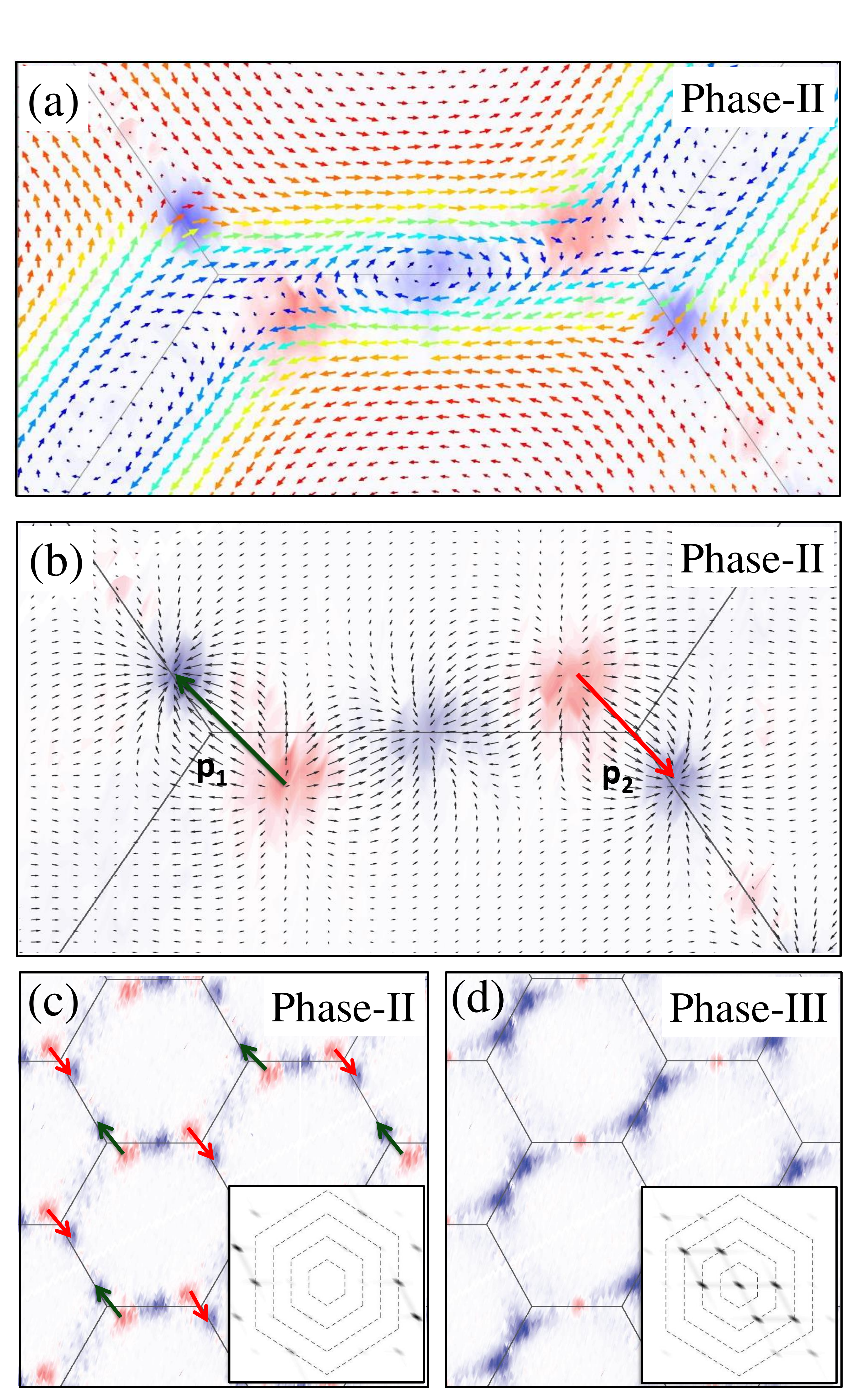}
\caption{{\bf Antiferroelectric Phase-II.}. (a-b) Zoomed-in view of the topological charge density in blue to red color map is plotted in the background, with arrows denoting spins ${\bf s}$ in (a), and emergent electric field ${\bf E}$ vector in (b). The spin texture clarifies the formation of topological charge at the Moir\'e lattice side, and a pair of  antiskyrmions with opposite charge centers near the Moir\'e lattice corners. The electric field lines in (b) confirm the formation of topological dipole moment (long arrows) between the antiskymion pairs. (c-d) Real space view of the topological charges in many Moir\'e unit cells for Phase-II and Phase-III, respectively. (c) We clearly observe the N\'eel analog of the ordering of the topological dipole moments, giving an antiferroelectric phase. (d) As we move from Phase-II to Phase-III, the antiskyrmion pairs are annihilated and integer topological charges become confined by a compact domain wall on different lattice sides. {\it Inset:} The structure factor of the corresponding topological charge density, showing no charge ordering in Phase-II, as opposed to Bragg peaks in Phase-III. 
}
\label{fig4}
\end{figure}

\subsection{Topological antiferroelectric phase}
Phase-II is very intriguing and novel, and requires separate discussions. The naive spin-texture of this phase (Fig.~\ref{fig1}(c)) is reminiscence of a spin spiral phase. However, unlike in the other trivial spiral phases, here several new types of topological charge centers are formed as shown in Fig.~\ref{fig3}(l) and \ref{fig4}(a). Firstly, we observe that the polarity density has a streamline flow diagram with 1D domain wall, see Fig.~\ref{fig3}(b). But it fails to commence a compact geometry to produce full skyrmion charge centers. However, there exists five sharp charge centers (three inside the $n_z=-1$ (blue) region and two outside). This structure periodically repeats in a smectic pattern. These charge centers have different origins from the previous three skyrmion phases with compact polarity density, and result from splitting of the vorticity by streamline flow of polarity density.

A zoomed-in view of the spin texture on top of the topological charge density, as shown in Fig.~\ref{fig4}(a), unravels the mechanism of these charge centers. The charge center at the middle (blue-colored charge density) is a `meron'-like structure, but with a fractional charge of $Q\sim 2/9\pm 0.025$. The other four charge centers form in pairs with opposite sign of charges at the Moir\'e zone corners. The corresponding spin textures reveal that they are {\it antiskyrmions},\cite{Antiskyrmion,Mn2RhSnParkin} with fractional charges $Q\sim \pm 2/9\pm 0.025$. We note that although there are five distinct meron like charge centers with $\pm 2/9$, they are shared between neighboring Moir\'e supercells. Apart from these concentrated point charges, another +2/9 (red) is spread over the Moir\'e cell which is not clearly visible in the colormap of Fig.~\ref{fig4}(a),(b). Hence, the total charge (point-charge and background-charge) within a given Moire supercell vanishes.

Each pair of oppositely charged antiskyrmions acts as a topological electric dipole, sitting at each lattice site of the Moir\'e lattice. The dipole moment is estimated to be $(0.05\pm 0.005)a$, where $a$ is the Moir\'e lattice constant. More interestingly, the dipoles are aligned {\it anti-ferroelectrically} between the nearest neighbors. This gives a N\'eel-like ordering of the topological dipole moments, in close analogy to the N\'eel order of magnetic moments in a honeycomb lattice.

Comparisons of the spin-textures between Phase-II (anti-ferroelectric) and Phase-III (skyrmions) throw light on the phase transition between them. The phase transition occurs when the antiskyrmion pairs coalesce. This also results in the closing of the streamline flow of the polarity density to form a compact domain wall (see Fig.~\ref{fig3}(c)). Hence a topological winding number description becomes appropriate in Phase-III. 

\subsection{Electromagnetic duality} To visualize the formation of the dipole moment in Phase-II, we write down a gauge-dual theory, and calculate the topological electric field lines. In an electronic quantum Hall analog, we know that a physical charge is attached to a magnetic flux via the Chern-Simons coupling $-$ also known as the Str\'eda formula.\cite{Streda} Using this analogy, we affix an emergent gauge field $a_{\mu}=(\phi,{\bf a})$ with the topological charge density $\rho$ as $\mathcal{J}_{\mu} = \epsilon_{\mu\nu\tau}\mathcal{F}_{\nu\tau}$. Here $\mathcal{F}_{\nu\tau}=\partial_{\nu}a_{\tau}$ is the corresponding curvature field tensor (an emergent electro-magnetic field, but not the same ${\bf B}$ field seen by local spins). The emergent `electric field' is read as $\mathbf{E}= -{\bf \nabla} \phi-\partial_t {\bf a}$. This `electric field' follows the Gauss' law, and acquires distinct spatial dependence according to the topological charge distributions. Using the Gauss law, the electric field lines can be found numerically from $\int {\bf E}.d{\bf S} = \rho$. The electric field lines shown by black arrows in Fig.~\ref{fig4}(b) confirm the existence of the dipole moment and their antiferroelectric ordering. We extend the calculations of electric field lines to all the other phases, and find that the the field line configurations are consistent with the ring-, rod-, and point-like charge centers as obtained in Phase-I, Phase-II, and Phase-V, respectively (see Fig.~\ref{figS5}).

\subsection{Topological correlation function}
In what follows, the topological dipole moment indeed provides a measure of the order parameter in Phase-II: it continuously disappears in going from Phase-II to Phase-III. To elucidate this further, we compare larger views of the Moir\'e lattice in Phase-II and Phase-III in Figs.~\ref{fig4}(c) and \ref{fig4}(d). We also calculate the topological charge susceptibility as $\chi_{c}({\bf r}) = \int d^2{\bf r}'\rho({\bf r'})\rho({\bf r+r'})$. The corresponding Fourier transformation gives the structure factor $S_{c}({\bf q})$, which  mimics the x-ray diffraction (XRD) pattern as seen by transmission electron microscopy (TEM), and may be indirectly probed by Lorentz TEM.\cite{Tokura10} The corresponding structure factors are plotted in the {\it inset} figures to Figs. \ref{fig4}(c) and \ref{fig4}(d). As expected, we observe Bragg peaks in Phase-III as the topological charges form a triangular lattice. The lattice also features a broken spatial rotation symmetry, and gives a smectic-like skyrmion lattice. In phase-II the charge centers do not exhibit any Bragg peak up to the third Brillouin zone. As the appearance of the new Bragg like peaks at the antiferromagnetic wavevector in the spin-spin correlation function, the antiferroelectric phase of the topological dipole moment may also be visualized in the dipole-dipole correlation function.

\section{Discussions and Conclusions}
So far, our discussions were primarily devoted to delineate the mechanism of skyrmion charges and antiskyrmion dipoles. Although not our primary focus here, it is worthwhile to outline few possible mechanism to destroy topological configurations, and the corresponding phase transitions. We discussed that the size of the compact domain wall ($r_{\rm d}$) as it reduces to zero or expands all the way to the Moir\'e lattice vector as a function of $J_{\perp}/J_{D}$ and the bilayer thickness $d$, it destroys the skyrmion configuration. Moreover,  energetic makes another dominant factor to destabilizing the saddle-point minima of skyrmions.
Once fluctuations are included, the topological charge centers oscillate, creating `phonon' like excitations, which melt the skyrmion crystals.\cite{MeltSkyrmion} We have studied the short range (nearest-neighbor) topological charge correlation function, and find that it exhibits a similar divergence behavior at all phase transition points (see Appendix C). In addition, we observe that the phase transition from Phase-II to Phase-III occurs via the coalescence of the  antiskyrmion pairs and vanishing dipole moment. This is reminiscence of the Kosterlitz-Thouless (KT) like transition, but generalized to the  $O(3)$ field.

Can we probe the topological electric field, dipole moment and the KT transition of the topological charge of the skyrmions? The electric field is the mediator of the force between two charge particles. Since a skyrmion charge is a topological charge, it  cannot be destroyed without deforming all the spins in a skyrmion. This prohibits two skyrmions of the same charge to come close to each other $-$ as if they experience a Coulomb repulsion between them. This phenomenon can be associated with an electric field. Much like how we measure an electric field by adding a test charge, here one can think of adding a test skyrmion in a skyrmion background, and study its dynamics. It will be found that the test skyrmion of same (opposite) charge will be repelled (attracted) from the skyrmion background. Similarly, when two skyrmions/antiskyrmions of opposite charges are spatially separated as in Phase-II, one can associate a dipole moment in the usual way. With tuning, the two opposite charges can either annihilate each other or the two charges can become unbound from each other. The second phenomena is analogous to the KT transition as seen in the vortex case. In the case of the KT transition in vortices, pairs of vortices of opposite charges are energetically favorable at low-temperature, which forms dipole moment and the material behaves as a dielectric. With increasing temperature, the vortex pairs split and the vortex charges become unbound, giving a plasma like phase. The phase transition between them is denoted by the KT transition. In our present case, we can speculate that a similar splitting of the dipole may occur with increasing temperature in Phase-II, giving us a unique opportunity to explore a possible KT transition of skyrmions.

Finally, the spin-spin correlation also plays an important role. We have calculated the  transverse and longitudinal spin-spin correlation functions in the static limit. The transverse component does not have any $q\sim 0$ mode, and remains nearly unchanged across all the phases. The longitudinal susceptibility shows Bragg peaks in four phases, but not in Phase-I. Because, Phase-I is non magnetic. These additional results and discussions are given in Appendix.

One may wonder how sensitive is the the phase diagram to DMI and SOC terms. We have checked that DMI brings in very little change to the spin configurations, and their topological properties are robust as long as the DMI strength is considerably weaker than $J_{\perp}$, $J_{\rm D}$. With an eye to synthesize twisted bilayers of VdW magnets,\cite{CrI3,Chalcogenides} is it known that the spins are local in nature, and the materials are charge insulators. Thus the SOC does not play an important role to the skyrmion configurations.

\appendix

\section{Effective inter-layer magnetic field}\label{Sec:AppA}

\begin{figure}[t]
\includegraphics[width=85mm]{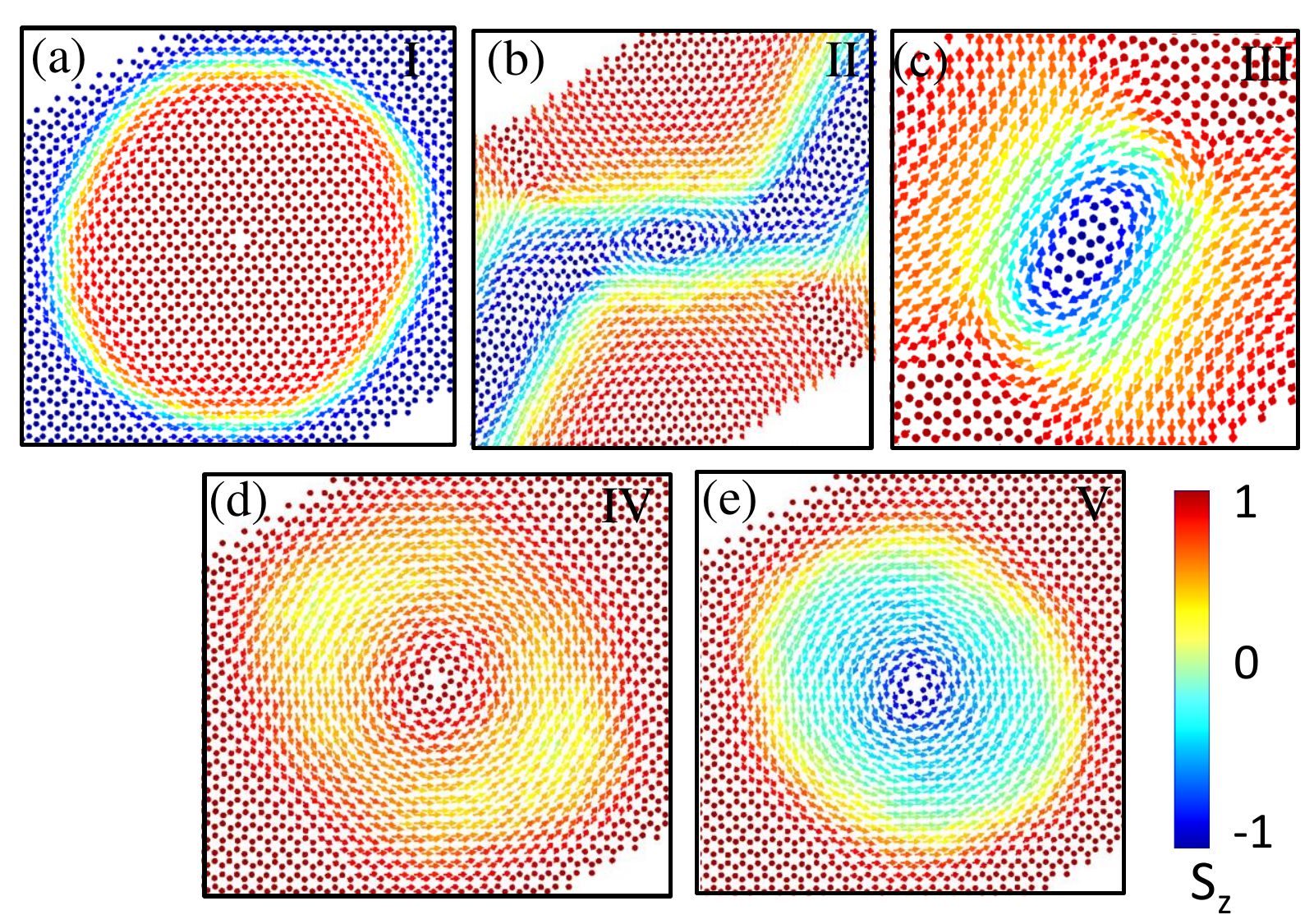}
\caption{Zoomed views of the spin structure in five different phases.}
\label{figS6}
\end{figure}

\begin{figure}[t]
\includegraphics[width=80mm]{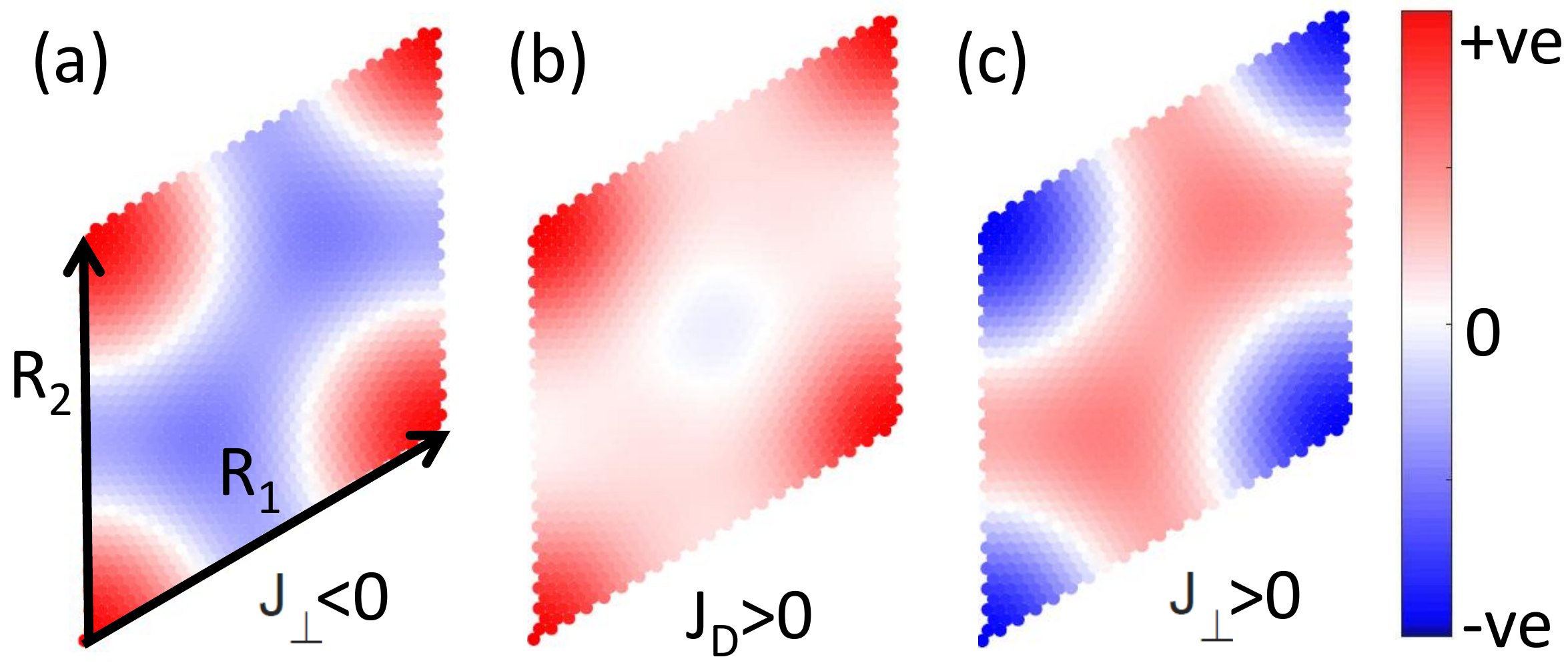}
\caption{Effective magnetic fields along the $z$-direction in the Moir\'e unit cell. (a) Field due to the inter-layer exchange interaction ($B_{ex}$) for $J_{\perp}<0$ and (c) for $J_{\perp}>0$. (b) Field due to the inter-layer dipole interaction ($B_{D}$) for $J_{D}>0$. ${\bf {R}}_{1}$ and ${\bf {R}}_{2}$ denote the two translation lattice vectors.}
\label{figS1}
\end{figure}

As mentioned in the main text, the effect of $H_{\rm ex}$ and $H_{\rm D}$ can be described by effective magnetic field ${\bf B}({\bf r}_{i})$ coming from exchange (${\bf B}_{\rm ex}({\bf r}_{i})$) and dipole (${\bf B}_{\rm D}({\bf r}_{i})$). Taking all the bottom layer's spin {\bf S} to be along the $+z$-direction, we can explicitly write the field components as
\begin{eqnarray}
{\bf B}_{\rm ex}({\bf r}_{i})&=& J_{\perp}\sum_{a=1,2,3}\cos({\bf G}_{a}.{\bf r}_{i}) \hat{{\bf z}},
\label{Eq:Bex} \\
{\bf B}_{\rm D}({\bf r}_{i})&=& \sum_{j}\frac{J_{D}}{r_{ij}^3}\left[ 1-\frac{3d^2}{r_{ij}^2} \right]\hat{{\bf z}} \nonumber \\
&&-\frac{3J_{D}d}{r_{ij}^5}\left[(x_i-x_j)\hat{{\bf x}}+(y_i-y_j)\hat{{\bf y}}\right].
\label{Eq:BD}
\end{eqnarray}
The index $i$ denotes a top layer spin. The sum over $j$ denotes a sum over bottom layer spins and is restricted up to $r_{\rm cut}=N_{\rm cut} a_0$. Due to the higher power of $r_{ij}$ in the denominator, it is easy to check that the summation converges very rapidly ($\sim N_{\rm cut} < 20$), much before the Moire supercell lattice vector $R_{1,2}\sim 35a_0$. ${\bf G}_{a}$ denotes three lowest Moir\'e reciprocal lattice vectors, $d$ is the inter-layer distance, and $x_i$, $y_i$ are $x$ and $y$ coordinates of $i^{\rm th}$ spin. ${\bf B}_{\rm ex}$ is plotted in fig.~\ref{figS1}(a) and (c) for $J_{\perp}<0$ (ferromagnetic) and $J_{\perp}>0$ (antiferromagnetic) respectively. 

Clearly, a compact domain wall forms at the nodal contour of $s_{z}({\bf r})$ where the total magnetic field roughly vanishes, i.e., $B_z({\bf r}_{\rm d})=0$. (This approximation works better where the in-plane spin exchange $J_{||} << |B|$, so that the second term in Eq.~\eqref{Eq:Hamfinal} dominates). Then the condition simplifies to
\begin{equation}
\frac{J_{\perp}}{J_{\rm D}}\beta_{ex}({\bf r}_{d}) + \beta_{D}({\bf r}_{d})=0.
\label{Eq:B_ratio}
\end{equation}
where ${\bf r}_d$ is the locii of the domain wall boundary and $\beta_{ex}$ and $\beta_{D}$ are given by
\begin{eqnarray}
\beta_{ex}({\bf r}_{d})&=&\sum_{a}\cos({\bf G}_{a}.{\bf r}_{d}) \\
\beta_{D}({\bf r}_{d})&=& \sum_{j}\frac{1}{|{\bf r}_d-{\bf r}_{j}|^3}\left[ 1-\frac{3d^2}{|{\bf r}_d-{\bf r}_{j}|^2} \right].
\end{eqnarray}
It is not easy to find an analytical expression for this nodal contour from Eqs.~\eqref{Eq:Bex}, and ~\eqref{Eq:BD}. 
But it's clear that the value of ${\bf r}_d$ depends on the $J_{\perp}/J_{\rm D}$ ratio and the bilayer thickness $d$ for a given Moir\'e lattice denoted by ${\bf G}$. The equation of the nodal line can be a circle or an ellipse depending on these parameters. As we see form the numerical simulation, the domain wall is circular for large values of $J_{\perp}$, while it takes an elliptical form in Phase-III for small values of $J_{\perp}$.

Again in the limit of $J_{||} << |B|$, a phase transition is defined at $J_{\perp}/J_{\rm D}=-\beta_{D}({\bf r}_{d})/\beta_{ex}({\bf r}_{d})$ (from Eq.~\eqref{Eq:B_ratio}) for different values of ${\bf r}_{d}$. In phase I, the radius of circular patch increases as $J_{\perp} \rightarrow 0$, and at the transition point to phase II two circular regions merge together at ${\bf r}_d={\bf R}_1/2$. On the other hand, ${\bf r}_d$ decreases in phase V as $J_{\perp} \rightarrow 0$, which implies ${\bf r}_d=0$ at the transition from the phase V to phase IV. Similarly transitions from phase II to III and III to IV are given by ${\bf r}_d={\bf R}_2/2$ and ${\bf r}_d=({\bf R}_1+{\bf R}_2)/2$ respectively. By numerically evaluating $\beta_{ex}$ and $\beta_{D}$ at various ${\bf r_d}$, we extract the approximate critical value of the ratio $J_{\perp}/J_{D}$ for different transitions which are listed in the table below:
\\
\\
\begin{tabular}{| p{3.5cm} | p{2.5cm} | p{1.5cm} |}
    \hline
     Transitions & ${\bf r}_d$ & $J_{\perp}/J_{D}$ \\ \hline \hline
    Phase I to phase II & ${\bf R}_1/2$ & -0.57 \\ \hline
    Phase II to phase III & ${\bf R}_2/2$ & -0.4 \\ \hline
    Phase III to phase IV & $({\bf R}_1+{\bf R}_2)/2$ & 0.3  \\ \hline
    Phase IV to phase V & ${\bf 0}$ & 1.6 \\
    \hline
     \end{tabular}
\\
\\

A direct comparison of transition from phase ‘I to II’ and phase ‘IV to V’ reveals that $|J_{\perp}/J_D|$ is larger on the positive side ($1.6$ on positive side and $-0.57$ on negative side). This can be understood from Fig.~6. At the corner of the Moir\'e supercell $B_{\rm ex}$ and $B_{\rm D}$ have the same sign in the $J_{\perp}<0$ region. So their effects add up. This is also responsible for a thin domain wall of magnetic bubbles in phase I. In the positive $J_{\perp}$ region, however, $B_{\rm ex}$ and $B_{\rm D}$ have opposite sign. So a larger value of $J_{\perp}$ is needed to overcome the effect $J_{\bf D}$ in positive side. This leads to a wider domain wall in phase V. We should distinguish the effects of the dipole-dipole interaction $H_{\rm D}$ of the present study compared to other studies. In earlier studies,\cite{SkyrmionDDI} such an interaction is involved for the same intra-layer spin (between ${\bf s}$ and ${\bf s}$ variables) which tends to produce magnetic bubble phases.\cite{SkyrmionReview1} On the other hand, here the dipole interaction is between the inter-layer (${\bf s}$ and ${\bf S}$), and is detrimental to the bubble or Bloch skyrmion phases, while promote the streamline flow of the polarity density.

\section{Detail of Monte Carlo method}\label{Sec:AppB}

As mentioned in the main text, the Hamiltonian for the top layer can be written as
\begin{eqnarray}
H=H_{1}+\sum_{i}{\bf {B}}({\bf {r}}_{i}).{\bf {s}}_{i}.
\end{eqnarray}
Here $H_{1}$ is the intra-layer Hamiltonian and ${\bf {B}}({\bf {r}}_{i})$ is the effective magnetic field due to inter-layer interaction. The lattice site ${\bf {r_{i}}}$ spans the Moir\'e unit cell. In our calculation at the commensurate angle $1.61^{0}$ we have 2522 basis site per unit cell. We set all the classical spins to have unit length ($S_{i}=1$) so that all the spins can be specified with two parameters $S_z$ and $\phi$ where $\phi$ is the angle that the component of spin on the $xy$-plane ($s^{xy}$) makes with the $x$ axis. From these two parameters, we can extract all the three components of the spins as
\begin{eqnarray}
s^{xy}_{i}=\sqrt{1-(s^{z}_{i}})^{2}, \nonumber \\
s^{x}=s^{xy}\cos(\phi), \nonumber \\
s^{y}=s^{xy}\sin(\phi).
\end{eqnarray}
We initialize the simulation with all spins pointing upward i.e., $S^{z}_{i}=1$, and then the next spin is chosen randomly from all the lattice points. The update algorithm for that spin is given by
\begin{eqnarray}
s^{z}_{i}=s^{z}_{i}+\gamma ds^{z}, \nonumber \\
\phi_{i}=\phi_{i}+\gamma d\phi.
\end{eqnarray}
And if $|s^{z}_{i}|>1$ then
\begin{eqnarray}
s^{z}_{i}=2 \mp (s^{z}_{i}+\gamma ds^{z}),\nonumber \\
\phi_{i}=\phi_{i}+\gamma d\phi+\pi.
\end{eqnarray}
Here the $\pm$ signs are for $s^{z}_{i}>1$ and $s^{z}_{i}<-1$, respectively. $\gamma$ is a random number between 1 and -1, and $ds^{z}$ and $d\phi$ are ranges of $s^{z}$ and $\phi$.

At each Monte Carlo step we calculate the local Hamiltonian
\begin{eqnarray}
H({\bf r}_{i})_{loc}&=&H_{1}({\bf r}_{i})-\sum_{i}{\bf {B}}({\bf {r}}_{i}).{\bf {s}}_{i}, \\
H_{1}({\bf r}_{i})&=&\sum_{<i,j>}J_{||}{\bf {s}}_{i}.{\bf {s}}_{j}.
\end{eqnarray}
and each configuration is accepted with a Boltzmann probability $e^{[H({\bf r}_{i})_{\rm loc}^{\rm new}-H({\bf r}_{i})_{\rm loc}^{\rm old}]/k_{B}T}$ at temperature $T$. To find the low temperature Monte Carlo ground state we perform temperature annealing as well as parameter annealing, and check for convergence of the total ground state energy. We use $dS^{z}=0.4$ and $d\phi=0.4\pi$. At low temperature we decrease the value of $ds^{z}(=0.1)$ and $d\phi(=0.1\pi)$ to increase the acceptance ratio in Monte Carlo steps. We equilibriate the system with $10^8$ steps and take the low energy configuration for another $10^8$ steps.

\section{Correlation functions and structure factors} 

\begin{figure}[t]
\includegraphics[scale=0.3]{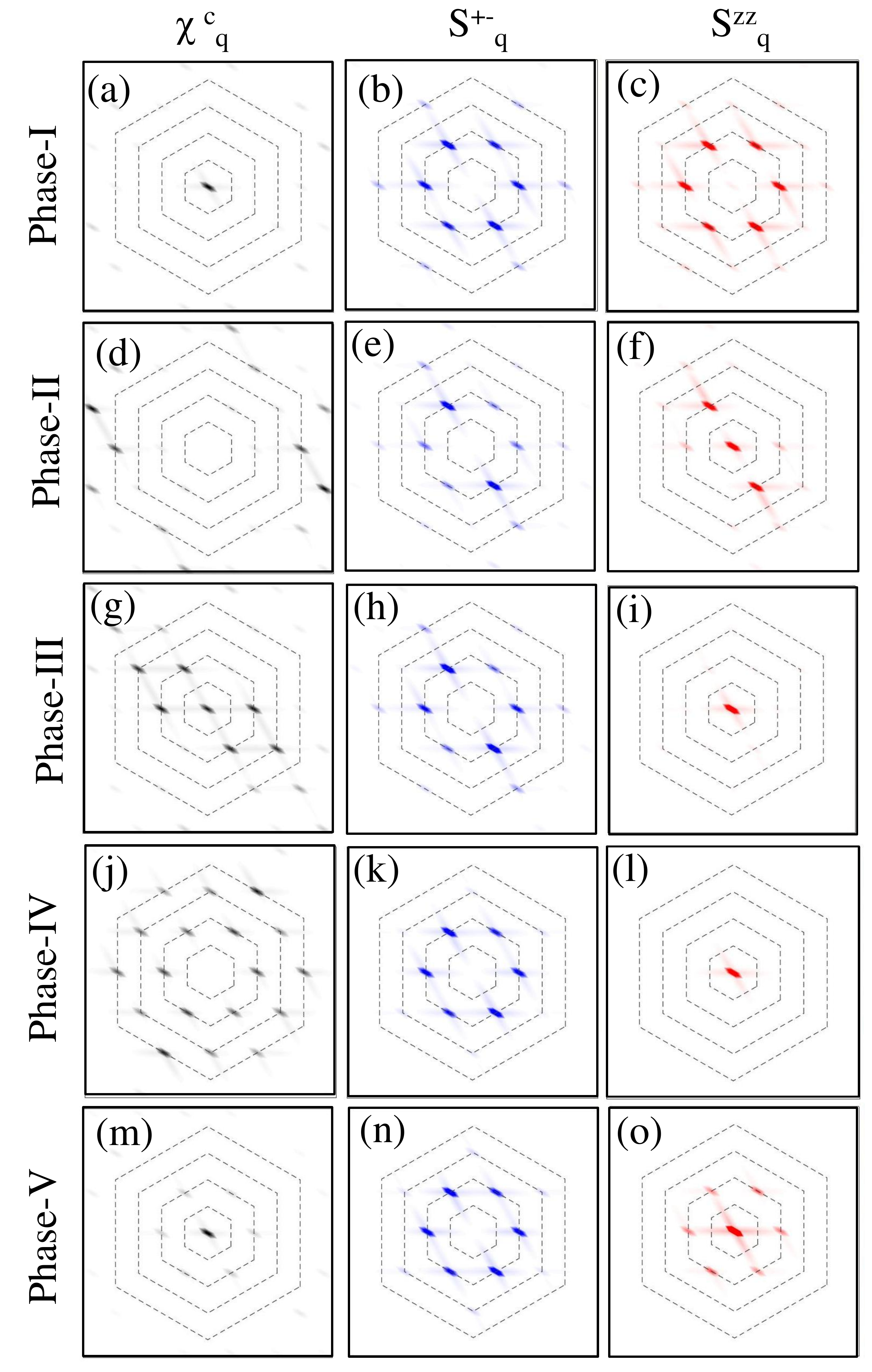}
\caption{Structure factors. (Left) Left row gives the structure factor of the topological charge density (Fourier transformation of the charge-charge correlation function). (Middle) Transverse spin-spin correlation functions. (Right) Longitudinal correlation function in the momentum space. In all three cases, we give various quadrants of the Brillouin zone with dashed lines. 
}
\label{figS2}
\end{figure}

From Fig.~2, it is evident that there are finite topological charge distributions even in the trivial topological phases where the polarity density $n_z$ fails to create domain walls. Therefore, to obtain a microscopic nature of how the quantum fluctuations make it possible to form a topological configurations in Phase-I, III and V, we investigate the following correlation functions: Topological charge susceptibility 
\begin{eqnarray}
\chi_{c}({\bf r}) = \int d^2{\bf r}'\rho({\bf r'})\rho({\bf r+r'}).
\end{eqnarray}
The spin-non-flip (or skyrmion polarity) and spin-flip correlation functions
\begin{eqnarray}
\chi_{s}^{zz}({\bf r})& =& \int d^2{\bf r}'n_z({\bf r'})n_z({\bf r+r'}),\nonumber\\
\chi_{s}^{\pm}({\bf r})& =& \int d^2{\bf r}'n_+({\bf r'})n_-({\bf r+r'}),
\end{eqnarray}
where $n_{\pm}=n_x\pm i n_y$. The Fourier transformation of the correlation function gives the structure factor $S_{c/s}({\bf q})$.  For the topological charge density, the topological structure factor mimics the XRD or TEM maps, except that this is not directly measurable and indirectly it can be mapped by Larentz TEM. The two spin structure factors are however measurable via  small-angle neutron scattering (SANS) experiments. 

First we notice that in the non-trivial phases, the topological structure factor shows distinct Bragg peaks at ${\bf q}\rightarrow 0$, signifying the transnational symmetry breaking and formation of skyrmion lattice. (When the dynamics is added, this peak will disperse away from ${\bf q}=0$  point as an typical acoustic Goldstone mode). In the trivial phase (Phase-II and Phase-IV), there is no Bragg peaks, but some weak intensity at higher Brillouin zone of the original Moir\'e supercell) which are often observed in dilute gas or liquid phases. 

The spin-non-flip and spin slip components of the spin-structure factors have the usual behavior in a skryrmions structure, while the stark difference between the Phase-I and Phase-IV for the $\chi_s^{zz}$ should be noted. In the case of a ring charge (Phase-I) there is no ${\bf q}\rightarrow 0$ mode for the polarity density since here we have a sharp domain wall boundary between the up and down spin states. For all other cases, the  ${\bf q}\rightarrow 0$ mode exists, suggesting a finite value of the magnetic moment in these cases on the top layer. On the other, in Phase-II, the total magnetization vanishes in a Moir\'e unit cell, however, if one define N\'eel magnetization between the up spin domain and down-spin domain, we have an antiferromagnetic like domain wall ordering. In all cases, the spin-flip structure factor is similar, while an additional spatial rotational symmetry breaking is observed in Phases II and III as expected. We find that that all the peaks are present in the second Brillouin zone which is due to the fact that the Moire cell here forms an honeycomb lattice which has two sublattices.

\begin{figure}[t]
\includegraphics[width=80mm]{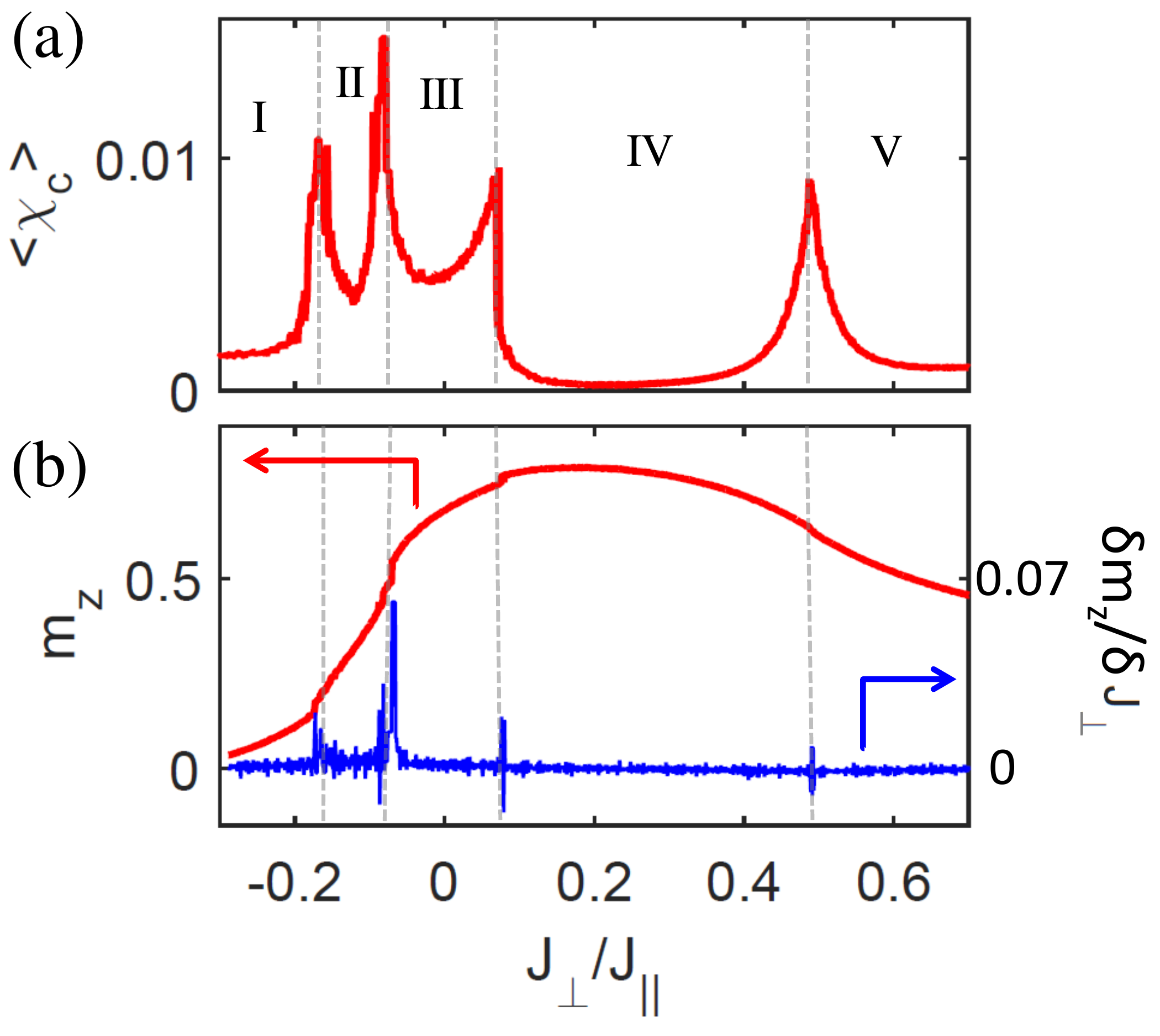}
\caption{plot of (a) Average $\chi_c$ and (b) average magnetization ($m_z$) and derivative of $m_z$ as a function of interlayer exchange ($J_{\perp}/J_{||}$) at a fixed $J_D=0.3J_{||}$. Discontinuity in $m_z$ at the phase transition can be seen from the jumps in derivative of $m_z$ (blue curve in (b))}
\label{figS3}
\end{figure}

To further elucidate the phase transition, we analyze the topological correlation function at a fixed distance ($r_{i} - r_{j}$=constant):
\begin{eqnarray}
\chi_{c}=\frac{1}{N}\sum_{<i,j>}\rho(\bf {r}_{i})\rho(\bf {r}_{j}).
\end{eqnarray}
We plot the result as a function of inter-layer exchange ($J_{\perp}$) in Fig.~\ref{figS3}(a). This gives the short-range correlation of the topological charge density. 

We find the correlation has peaks at the phase transition points as shown in Fig.~\ref{figS3}(a). This result is consistent with jumps in the effective magnetic field $M_{z}=\left\langle (1/N)\sum_{i}S_{i}^{z} \right\rangle $ along $J_{\perp}$, as shown in Fig.~\ref{figS3}(b). Near the phase transition points, the landscape of effective magnetic field changes over the Moir\'e unit cell. As a result near the domain wall in each phase (except Phase IV where there is no domain wall) ${\bf B}$ field becomes small over a large area. Therefore, the Hamiltonian is mostly dominated by strong in-plane ferromagnetic exchange ($J_{||}$), and produces a strong nearest neighbour correlation.

We further calculate scalar chirality (not shown)$\chi_{sc}=\frac{1}{N}\left\langle \sum_{<i,j,k>}{\bf S}_{i}.\left( {\bf S}_{j} \times {\bf S}_{k}\right) \right\rangle $, where  $\left\langle i,j,k\right\rangle $ denotes three spins forming the smallest triangle in the honeycomb lattice. We find that the scalar chirality vanishes in Phase II and Phase IV, where the magnetization has a sharp jump at their phase boundary (Fig.~\ref{figS3}(b)).

\section{Dipole and quadrupole moments in Phase II}

\begin{figure}[t]
\includegraphics[width=85mm]{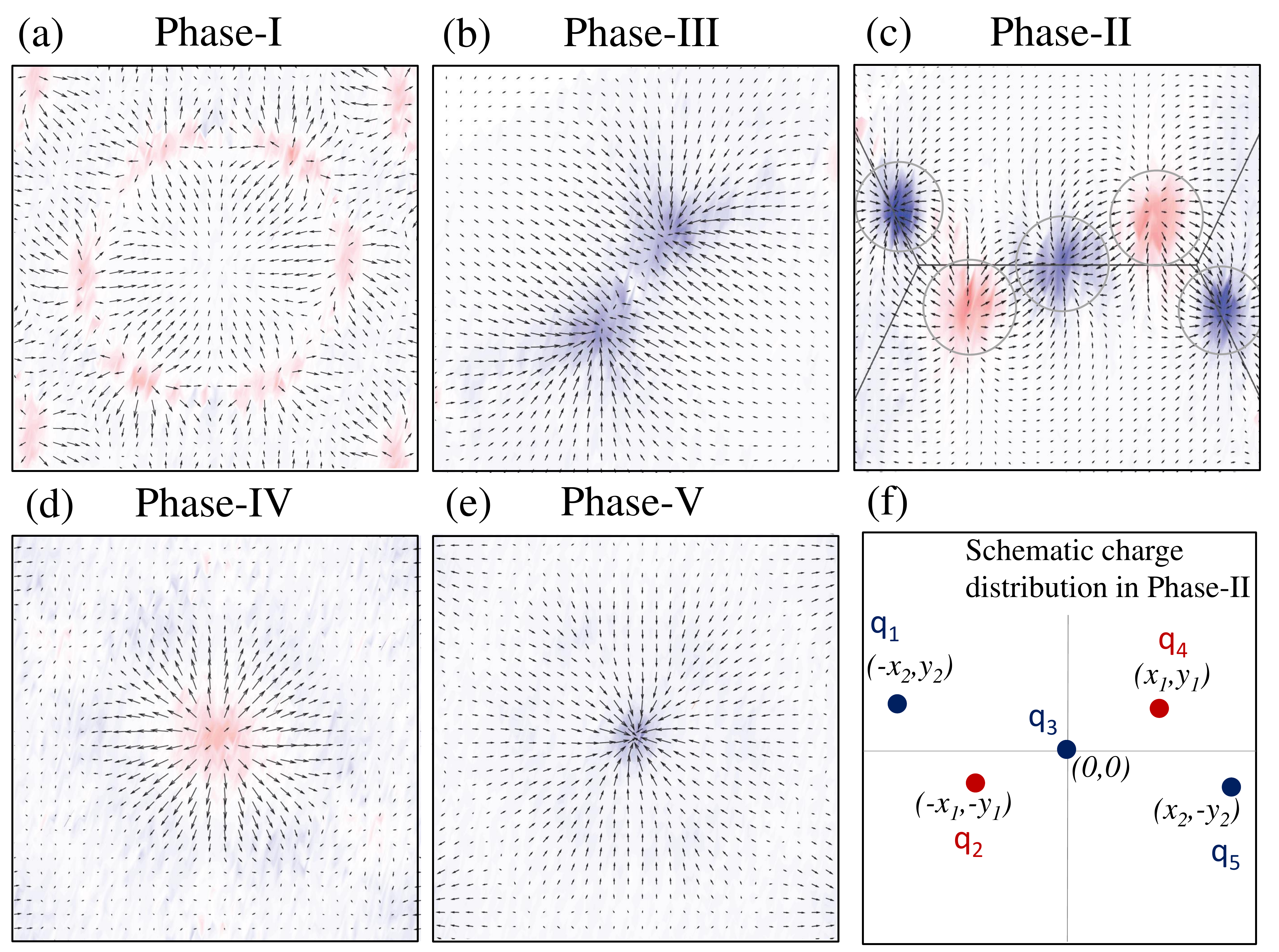}
\caption{(a-e)We have plotted the emergent topological electric field lines on top of the topological charge density in the vicinity of the charge centers (see Fig.~\ref{figS4} for larger view) in all five phases. In (f) we show the schematic distribution of point charges for Pase II with origin of the coordinates at the cahrge $q_3$.}
\label{figS5}
\end{figure}

Field lines for topological charge in Phase II is shown in Fig.~\ref{figS5}(c). The field lines indicate that the charge distribution can be approximated with point charges ($q_i$, $i=1,2,3,4,5$) as shown in Fig.~\ref{figS5}(f). As the total charge of this configuration is zero we investigate the higher moments (dipole and quadrupole) of this charge distribution. We calculate dipole moment for charge pair $(q_1,q_2)$ and $(q_4,q_5)$ and taking the origin of our co-ordinate at one of the charge center (see Fig.~\ref{figS5}(f)) we calculate the quadrupole moments which are given by
\begin{eqnarray}
Q &=& \left[
\begin{array}{ c c }
Q_{xx} & Q_{xy}\\
Q_{yx} & Q_{yy}
\end{array} \right] \ \
=\left[
\begin{array}{ c c }
\sum_i \rho_i x_i x_i & \sum_i \rho_i x_i y_i\\
\sum_i \rho_i y_i x_i & \sum_i \rho_i y_i y_i
\end{array} \right]
\end{eqnarray}
Values of the dipole and quadrupole moments for Phase II are given in the following table.

\begin{tabular}{| p{2.8cm} | p{2.4cm} | p{2.4cm} |}
    \hline
     quantities & $J_D=0.53$ & $J_D=1.0$ \\ \hline \hline
    $q_1$ & -0.23 & -0.23 \\ \hline
    $q_2$ & 0.22 & 0.22 \\ \hline
    $q_3$ & -0.25 & -0.24 \\ \hline
    $q_4$ & 0.22 & 0.24 \\ \hline
    $q_5$ & -0.20 & -0.22 \\ \hline
    $(x_1,y_1)$ in umit of moir\'e lattice vector R=35.5 & (0.20 , 0.07) & (0.20 , 0.07) \\ \hline
    $(x_2,-y_2)$ in umit of moir\'e lattice vector R=35.5 & (0.35 , -0.09) & (0.35 , -0.09) \\ \hline
    ${\bf p}_1$ & 0.033$\hat{\bf x}$ + 0.035$\hat{\bf y}$ & 0.033$\hat{\bf x}$ + 0.035$\hat{\bf y}$ \\ \hline
    ${\bf p}_2$ & -0.033$\hat{\bf x}$ - 0.035$\hat{\bf y}$ & -0.034$\hat{\bf x}$ - 0.037$\hat{\bf y}$ \\ \hline
    $Q_{xx}$ & -0.035 & -0.037 \\ \hline
    $Q_{yy}$ & -0.001 & -0.001 \\ \hline
    $Q_{xy}$ & 0.021 & 0.020 \\
    \hline
     \end{tabular}
  
We caution the readers that the estimate of the fractional charge is subjected to the integration contour around each charge centers, since there is no compact  of domain wall boundary here.

\begin{figure}[t]
\includegraphics[width=85mm]{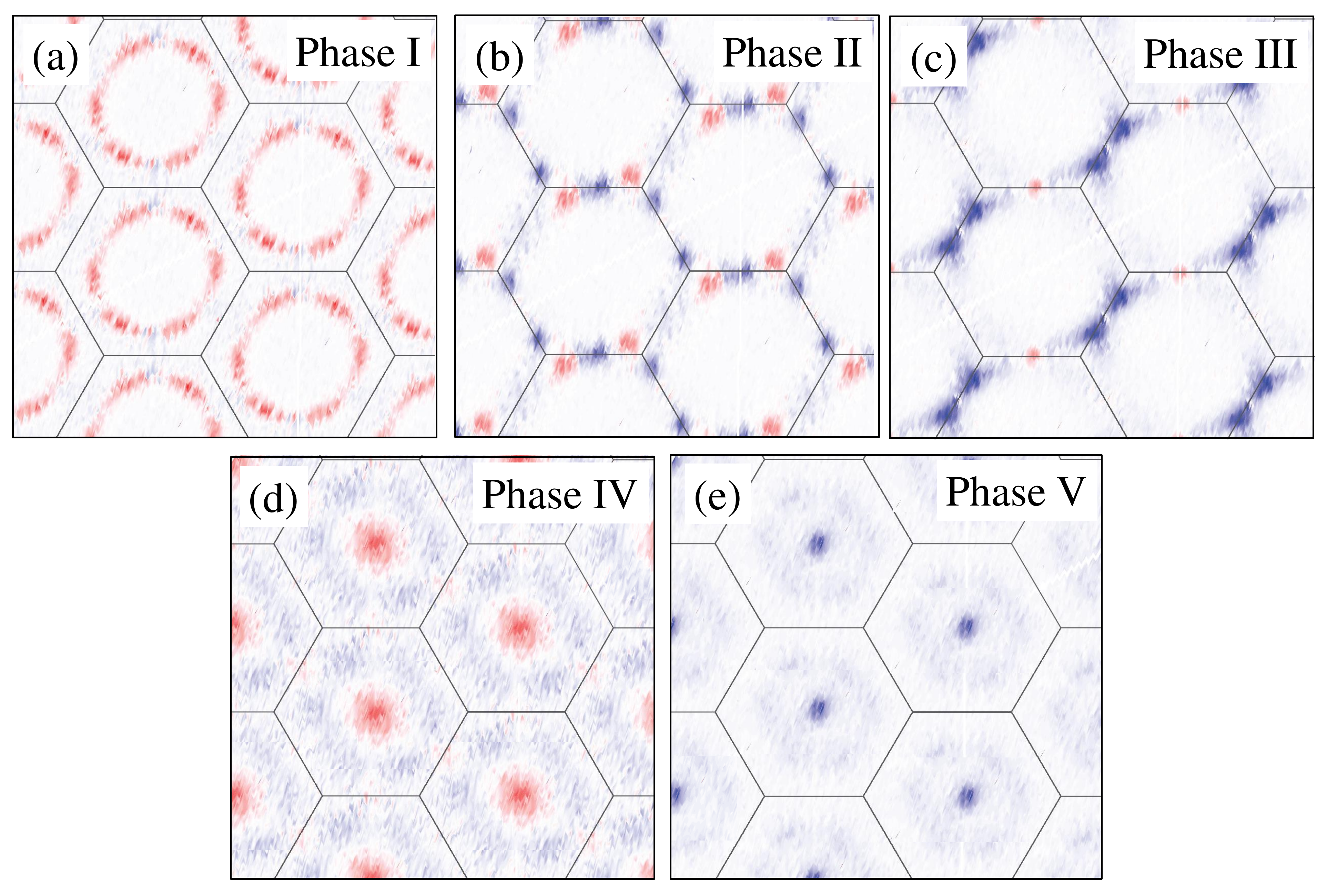}
\caption{Plots of topological charge density in many Moir'e unit cells.}
\label{figS4}
\end{figure}


\clearpage

\begin{thebibliography}{99}
\bibitem{Skyrme} T. H. R. Skyrme, A non-linear field theory. Proc. Royal Soc. Lond. A: Math. Phys. Eng. Sci. {\bf 260}, 127 (1961); T. H. R. Skyrme, A unified field theory of mesons and baryons. Nucl. Phys. {\bf 31}, 556 (1962).

\bibitem{SkyrmionGenealTheory} A. N. Bogdanov and A. Hubert, Sov.Phys.JETP {\bf 68}, 101-103 (1989); J. Magn. Magn. Mater. {\bf138} , 255-269 (1994); 
U. K. Roessler, A. N. Bogdanov, and C. Pfleiderer, Nature {\bf 442} , 797-801 (2006).

\bibitem{SkyrmionReview1}N. Nagaosa and Y. Tokura, Nature Nanotechnology {\bf 8}, 899-911 (2013).

\bibitem{Skyrmionbook}J. H. Han, Skyrmions in
Condensed Matter, Springer Tracts in Modern Physics, Volume {\bf 278}

\bibitem{SkyrmionDDI}Y. S. Lin, J. Grundy, and E. A. Giess, Appl. Phys. Lett. {\bf 23}, 485–487 (1973);
A. P. Malozemoff, and J. C. Slonczewski, Magnetic Domain Walls in Bubble Materials 306–314 (Academic, 1979).; T. Garel, and S. Doniach, Phys. Rev. B {\bf 26}, 325–329 (1982);
T. Suzuki, J. Magn. Magn. Mater. {\bf 31–34}, 1009–1010 (1983);
B. Göbel, J. Henk, I. Mertig, Sci. Reports {\bf 9}, 9521 (2019).
M. Ezawa, Phys. Rev. Lett. {\bf 105}, 197202 (2010).

\bibitem{SkyrmionDMI}S. Rohart, and A. Thiaville, Phys. Rev. B {\bf 88}, 184422 (2013).

\bibitem{skDMI1}M. Hoffmann, B. Zimmermann, G. P. Muller, D. Schurhoff, N. S. Kiselev, C. Melcher, and S. Blugel. Nat. comm. {\bf 8}, 308 (2017).

\bibitem{skDMI2}E. Y. Vedmedenko, J. A. Arregi, P. Riego, A. Berger. Phys. Rev. Lett. {\bf 122}, 257 (2019)

\bibitem{skDMI3}S. Rohart and A. Thiaville, Phys. Rev. B Condens. Matter Mater. Phys. {\bf 88}, 184422 (2013).

\bibitem{skDMI4}B. Dup\'e, G. Bihlmayer, M. Böttcher, S. Blügel, and S. Heinze, Nat. Commun. {\bf 7}, 11779 (2016).

\bibitem{magnetic_frus1} T. Okubo, S. Chung and H. Kawamura, Phys. Rev. Lett. {\bf 108}, 017206 (2012).

\bibitem{magnetic_frus2} A. O. Leonov, and M. Mostovoy, Nat. Commun. {\bf 6}, 8275 (2015).

\bibitem{magnetic_frus3} C. D. Batista, S. Z. Lin, S. Hayami, and Y. Kamiya, Rep. Prog. Phys. {\bf 79}, 084504 (2016).

\bibitem{SkyrmionFrus}T. Okubo, S. Chung, and H. Kawamura, Multiple-q states and the skyrmion lattice of the triangular-lattice Heisenberg antiferromagnet under magnetic fields. Phys. Rev. Lett. {\bf 108}, 017206 (2012); 
A. Leonov, and M. and Mostovoy, Multiply periodic states and isolated skyrmions in an anisotropic frustrated magnet. Nat. Commun. {\bf 6}, 8275 (2015); S. Hayami, S. -Z. Lin, and Cristian D. Batista Phys. Rev. B {\bf 93}, 184413 (2016); C. D. Batista, S. -Z. Lin, S. Hayami, and K. Yoshitomo, Reports on Progress in Physics, {\bf 79}, 084504 (2016).

\bibitem{SkyrmionSOC}S. Banerjee, O. Erten, and M. randeria, Nature Physics {\bf 9}, 626-630 (2013).

\bibitem{magnetic_frus4} R. Ozawa, S. Hayami, and Y. Motome, Phys. Rev. Lett. {\bf 118}, 147205 (2017).

\bibitem{skyrmion_disorder} I. Gross, W. Akhtar, A. Hrabec, J. Sampaio, L. J. Martinez, S. Chouaieb, B. J. Shields, P. Maletinsky, A. Thiaville, S. Rohart, and V. Jacques, Phys. Rev. Mater. {\bf 2}, 024406 (2017).

\bibitem{SkyrmionQHI}S. L. Sondhi, A. Karlhede, S. A. Kivelson, and E. H. Rezayi, Phys. Rev. B {\bf 47}, 16419 (1993).

\bibitem{SkyrmionBEC}L. S. Leslie, A. Hansen, K. C. Wright, B. M. Deutsch, and N. P. Bigelow, Phys. Rev. Lett. {\bf 103}, 250401 (2009).

\bibitem{MnSi}S. M\"uhlbauer, B. Binz, F. Jonietz, C. Pfleiderer, A. Rosch, A. Neubauer, R. Georgii, and P. B\"oni, Science {\bf 323}, 915-919 (2009).

\bibitem{FeCoSi}W. M\"unzer, A. Neubauer, T. Adams, S. M\"uhlbauer, C. Franz, F. Jonietz, R. Georgii, P. B\"oni, B. Pedersen, M. Schmidt, A. Rosch, and C. Pfleiderer Phys Rev. B {\bf 81}, 041203(R) (2010).

\bibitem{Tokura10} X. Z. Yu, Y. Onose, N. Kanazawa, J. H. Park, J. H. Han, Y. Matsui, N. Nagaosa and Y. Tokura, Nature {\bf 465}, 901–904 (2010); D. Morikawa, K. Shibata, N. Kanazawa, X. Z. Yu, and Y. Tokura, Phys. Rev. B {\bf 88} 024408 (2013).

\bibitem{MildeKohlerSci13}P. Milde, D. K\"ohler, J. Seidel, L. M. Eng, A. Bauer, A. Chacon, J. Kindervater, S. M\"uhlbauer, C. Pfleiderer, S. Buhrandt, C. Sch\"utte, and A. Rosch, Science {\bf 340}, 1076-1080 (2013).

\bibitem{Cu2OSeO3}S. Seki, X. Z. Yu, S. Ishiwata, and Y. Tokura, Science {\bf 336}, 198-201 (2012).

\bibitem{CoZnMn}Y. Tokunaga, X. Z. Yu, J. S. White, H. M. Ronnow, D. Morikawa, Y. Taguchi, and Y. Tokura, Nat. Commun. {\bf 6}, 7638 (2015).

\bibitem{FeGe}X.Zhao, C. Jin, C. Wang, H. Du, J. Zang, M. Tian, R. Che, and Y. Zhang, Proc. Natl. Acad. Sci. {\bf 113}, 4918 (2016). 

\bibitem{heavymetals}S. Heinze, K. von Bergmann, M. Menzel, J. Brede, A. Kubetzka, R. Wiesendanger, G. Bihlmayer, S. Bl\"ugel, Nat. Phys. {\bf 7}, 713-718 (2011);

\bibitem{Kagome}Z. Hou, W. Ren, B. Ding, G. Xu, Y. Wang, B. Yang, Q. Zhang, Y. Zhang, E. Liu, F. Xu, W. Wang, G. Wu, X. -X. Zhang, B. Shen, Z. Zhang, Advanced Materials {\bf 29}, 1701144 (2017).

\bibitem{WisendangerReview} R. Wiesendanger, Nanoscale magnetic skyrmions in metallic films and multilayers: a new twist for spintronics. Nat. Rev. Mat. {\bf 1}, 16044 (2016).

\bibitem{Mn2RhSnParkin}A. K. Nayak, V. Kumar, T. Ma, P. Werner, E. Pippel, R. Sahoo, F. Damay, U. K. R\"u$\beta$ler, C. Felser and S. S. P. Parkin, Nature {\bf 548}, 561-566 (2017).

\bibitem{Antiskyrmion}W. Koshibae, and  N. Nagaosa, Nat. Commun., {\bf 7}, 10542 (2016);  M. Hoffmann, B. Zimmermann, G. P. M\"uller, D. Sch\"urhoff, N. S. Kiselev, C. Melcher, and S. Bl\"ugel  Nat. Commun. {\bf 8}, 308 (2017)

\bibitem{WritingSkyrmion}N. Romming, C. Hanneken, M. Menzel, J. E. Bickel, B. Wolter, K. von Bergmann, A. Kubetzka, and R. Wiesendanger, Science {\bf 341}, 636-639 (2013).

\bibitem{MobileSkyrmion}J. Sampaio, V. Cros, S. Rohart, A. Thiaville, and A. Fert, Nature Nanotechnology {\bf 8}, 839–844 (2013).

\bibitem{MochizukiNatMat}M. Mochizuki, X. Z. Yu, S. Seki, N. Kanazawa, W. Koshibae, J. Zang, M. Mostovoy, Y. Tokura, and N. Nagaosa, Nature Materials {\bf 13}, 241–246 (2014).

\bibitem{Hofman}W. Jiang, P. Upadhyaya, W. Zhang, G. Yu, M. B. Jungfleisch, F. Y. Fradin, J. E. Pearson, Y. Tserkovnyak, K. L. Wang, O. Heinonen, S. G. E. te Velthuis, and A. Hoffmann, Science {\bf 349}, 283-286 (2015).

\bibitem{SkyrmionReviewTech} A. Fert, N. Reyren, and V. Cros, Nature Reviews Materials, {\bf 2}, 17031 (2017); A. Fert, V. Cros, and J. Sampaio, Skyrmions on the track. Nat. Nanotechnol. {\bf 8}, 152-156 (2013).

\bibitem{RacetrackParkin}S. S. P. Parkin, M. Hayashi, and L. Thomas, Science {\bf 320}, 190–194 (2008).

\bibitem{Racetrack2} R. Tomasello, E. Martinez, R. Zivieri, L. Torres, M. Carpentieri, and G. Finocchio, Sci. Rep. {\bf 4}, 6784 (2014); P. F. Bessarab, G. P. M\"uller, I. S. Lobanov, F. N. Rybakov, N. S. Kiselev, H. J\'onsson, V. M. Uzdin, S. Bl\"ugel, L. Bergqvist, and A. Delin, Sci. Reports {\bf 8}, 3433 (2018).

\bibitem{FePS3}X. Wang, K. Du, Y. Y. F. Liu, P. Hu, J. Zhang, Q. Zhang, M. H. S. Owen, X. Lu, C. K. Gan, and P. Sengupta, 2D Materials {\bf 3}, 3, 031009 (2016).

\bibitem{FePS3a} J. U. Lee, S. Lee, J. H. Ryoo, S. Kang, T. Y. Kim, P. Kim, C. H. Park, J. G. Park, and H. Cheong, Nano Lett. {\bf 16}, 12, 7433-7438 (2016).

\bibitem{MPX} P. A. Joy, and S. Vasudevan, Phys. Rev. B {\bf 46}, 5425-5433 (1992).

\bibitem{MPXa} C. C. Mayorga-Martinez, Z. Sofer, D. Sedmidubsky, S. Huber, A. Y. S. Eng, and M. Pumera, ACS Appl. Mater. Interfaces {\bf 9}, 12563–12573 (2017).

\bibitem{FGT} Z. Fei, B. Huang, P. Malinowski, W. Wang, T. Song, J. Sanchez, W. Yao, D. Xiao, X. Zhu, A. F. May, W. Wu, D. H. Cobden, J. -H. Chu, and X. Xu, Nat.Materials, {\bf 17}, 778–782 (2018).

\bibitem{FGTa}H. J. Deiseroth, K. Aleksandrov, C. Reiner, L. Kienle, and R. K. Kreme, Eur. J. Inorg. Chem., 1561–1567 (2006).

\bibitem{MnX} M. Kan, S. Adhikari, and Q. Sun, Phys. Chem. Chem. Phys. {\bf 16}, 4990-4994 (2016).

\bibitem{MnXa} C. Ataca, andH. Sahin, and S. Stable, J. Phys. Chem. C {\bf 116}, 8983-8999 (2012).

\bibitem{VX} Y. Ma, Y. Dai, M. Guo, C. Niu, Y. Zhu, B. Huang, ACS Nano {\bf 6}, 1695-1701 (2013).

\bibitem{VXa} M. Bonilla, S. Kolekar, Y. Ma, H. C. Diaz, V. Kalappattil, R. Das, T. Eggers, H. R. Gutierrez, M. -H. Phan, and M. Batzill, Nat. Nanotechnology {\bf 13}, 289-293 (2018).

\bibitem{VXb} G. Duvjir, B. K. Choi, I. Jang, S. Ulstrup, S. Kang, T. T. Ly, S. Kim, Y. H. Choi, C. Jozwiak, A. Bostwick, E. Rotenberg, J. -G. Park, R. Sankar, K. -S. Kim, J. Kim, and Y. Jun Chang, Nano. Lett. {\bf 18}, 5432-5438 (2018).

\bibitem{CrX} C. Gong, L. Li, Z. Li, H. Ji, A. Stern, Y. Xia, T. Cao, W. Bao, C. Wang, Y. Wang, Z. Q. Qiu, R. J. Cava, S. G. Louie, J. Xia, X. Zhang, Nature {\bf 546}, 265-269 (2017).

\bibitem{CrXa} B. Huang, G. Clark, E. N.-Moratalla, D. R. Klein, R. Cheng, K. L. Seyler, D. Zhong, E. Schmidgall, M. A. McGuire, D. H. Cobden, W. Yao, D. Xiao, P. J.-Herrero, and X. Xu, Nature {\bf 546}, 270-273 (2017).

\bibitem{CrXbi} B. Huang, G. Clark, D. R. Klein, D. MacNeill, E. N. -Moratalla, K. L. Seyler, N. Wilson, M. A. McGuire, D. H. Cobden, D. Xiao, W. Yao, P. J. -Herrero, and X. Xu, Nat. Nanotechnology {\bf 13}, 544–548 (2018).

\bibitem{CrXbia} S. Jiang, L. Li, Z. Wang, K. F. Mak, and J. Shan, Nat. Nanotechnology {\bf 13}, 549–553 (2018).

\bibitem{CrI3_DFT}N. Sivadas, S. Okamoto, X. Xu, C. J. Fennie, D. Xiao, Nano Lett. {\bf 18}, 7658-7664 (2018).
\bibitem{Moire_magnets}K. Hejazi, Z. -X. Luo, and L. Balents, PNAS {\bf 117}, 10721-10726 (2020).

\bibitem{skTBG} T. B\"omerich, L. Heinen, and A. Rosch, Phys. Rev. B {\bf 102}, 100408(R) (2020).

\bibitem{Substrate}Q. Tong, F. Liu, J. Xiao, and W. Yao Nano Lett. {\bf 18}, 7194-7199 (2018).

\bibitem{skFerroDMI} M. Akram, and O. Erten, Phys. Rev. B {\bf 103}, L14040 (2021).

\bibitem{FGT_CoPd} M. Yang, Q. Li, R. V. Chopdekar, R. Dhall, J. Turner, J. D. Carlstr\"om, C. Ophus, C. Klewe, P. Shafer, A. T. N'Diaye, J. W. Choi, G. Chen, Y. Z. Yu, C. Hwang, F. Wang, and Z, Q. Qiu, Science Advances {bf 6}, eabb5157 (2020).

\bibitem{FGT_hBN} T.-E. Park, L. Peng, J. Liang, A. Hallal, F. S. Yasin, X. Zhang, S. J. Kim, K. M. Song, K. Kim, M. Weigand, G. Schuetz, S. Finizio, J. Raabe, K. Garcia, J. Xia, Y. Zhou, M. Ezawa, X. Liu, J. Chang, H. C. Koo, Y. D. Kim, M. Chshiev, A. Fert, H. Yang, X. Yu, and S. Woo, Phys. Rev. B {\bf 103}, 104410 (2021).

\bibitem{continum_model}T. B\"omerich, L. Heinen, and A. Rosch, arXiv:2004.13684v2;

\bibitem{CrI3}M. A. McGuire, H. Dixit, V. R. Cooper, and B. C. Sales, Chem. Mater. {\bf 27}, 612–620 (2015);
B. Huang, G. Clark, E. N. -Moratalla, D. R. Klein, R. Cheng, K. L. Seyler, D. Zhong, E. Schmidgall, M. A. McGuire, D. H. Cobden, W. Yao, D. Xiao, P. J. -Herrero, and X. Xu, Nature {\bf 546}, 270–273 (2017).

\bibitem{Chalcogenides}C. Gong, L. Li, Z. Li, H. Ji, A. Stern, Y. Xia, T. Cao, W. Bao, C. Wang, Y. Wang, Z. Q. Qiu, R. J. Cava, S. G. Louie, J. Xia, and X. Zhang, Nature {\bf 546}, 265–269 (2017);
M. Bonilla, S. Kolekar, Y. Ma, H. C. Diaz, V. Kalappattil, R. Das, T. Eggers, H. R. Gutierrez, M. -H. Phan, and M. Batzill, Nat. Nanotech. {\bf 13}, 289–293 (2018);
D. J. \'OHara, T. Zhu, A. H. Trout, A. S. Ahmed, Y. K. Luo, C. H. Lee, M. R. Brenner, S. Rajan, J. A. Gupta, D. W. McComb, and R. K. Kawakami, Nano Lett. {\bf 18}, 3125−3131 (2018).

\bibitem{DasRMP} A. Bansil, H. Lin, T. Das, Rev. Mod. Phys. {\bf 88}, 021004 (2016).

M. Z. Hasan, C. L. Kane, Rev. Mod. Phys. {\bf 82}, 3045 (2010).

X. L. Qi, S. C. Zhang, Rev. Mod. Phys. {\bf 83}, 83, 1057 (2011).

\bibitem{MagBubble}A. Malozemoff, and J. Slonczewski, Magnetic Domain Walls in Bubble Materials (Academic press, 1979); A. H. Eschenfelder, Magnetic bubble technology (Springer-Verlag Berlin Heidelberg New York, 1980).

\bibitem{DiracNodal}T. Bzdušek, Q. Wu, A. R\"uegg, M. Sigrist, and A. A. Soluyanov, Nature {\bf 538}, 75–78 (2016).

\bibitem{Streda}P. Streda, J. Phys. C: Solid State Phys., {\bf 15}, L717-L721 (1982).

\bibitem{MeltSkyrmion} P. Huang, {\it et al.} Nat. Nanotech. {\bf 15}, 761-767 (2020).

\end{thebibliography}
\end{document}